\documentclass[11pt, preprint]{aastex63}

\submitjournal{ApJ}

\shorttitle{Planetary obliquity from frequency modulation}
\shortauthors{Nakagawa et al.}
\begin{document}

\title{Obliquity of an Earth-like planet from frequency modulation of
  its direct imaged lightcurve: mock analysis from general circulation
  model simulation }

\correspondingauthor{Yasushi Suto}
\email{suto@phys.s.u-tokyo.ac.jp}

\author[0000-0002-1004-685X]{Yuta Nakagawa}
\affiliation{Department of Physics, The University of Tokyo, Tokyo
  113-0033, Japan}

\author[0000-0001-9032-5826]{Takanori Kodama}
\affiliation{Laboratoire d'astrophysique de Bordeaux, Universit\'{e}
  de Bordeaux, CNRS, All\'{e}e Geoffroy Saint-Hilaire, 33615 Pessac,
  France}

\author[0000-0001-7490-4676]{Masaki Ishiwatari}
\affiliation{Division of Earth and Planetary Sciences, Hokkaido
  University, Sapporo 060-0810, Japan}
\affiliation{Center for Planetary Science, Kobe University, Kobe
  650-0047, Japan}

\author[0000-0003-3309-9134]{Hajime Kawahara}
\affiliation{Department of Earth and Planetary Science, The University
  of Tokyo, Tokyo 113-0033, Japan}
\affiliation{Research Center for the Early Universe, School of
  Science, The University of Tokyo, Tokyo 113-0033, Japan}

\author[0000-0002-4858-7598]{Yasushi Suto}
\affiliation{Department of Physics, The University of Tokyo, Tokyo
  113-0033, Japan}
\affiliation{Research Center for the Early Universe, School of
  Science, The University of Tokyo, Tokyo 113-0033, Japan}

\author[0000-0003-4060-7379]{Yoshiyuki O. Takahashi}
\affiliation{Center for Planetary Science, Kobe University, Kobe
  650-0047, Japan}
\affiliation{Department of Planetology, Kobe University, Kobe
  657-8501, Japan}

\author[0000-0002-3821-6881]{George L. Hashimoto}
\affiliation{Department of Earth Sciences, Okayama University, Okayama
  700-8530, Japan}

\author[0000-0002-6757-8064]{Kiyoshi Kuramoto}
\affiliation{Division of Earth and Planetary Sciences, Hokkaido
  University, Sapporo 060-0810, Japan} \affiliation{Center for
  Planetary Science, Kobe University, Kobe 650-0047, Japan}

\author[0000-0002-0333-2364]{Kensuke Nakajima}
\affiliation{Department of Earth and Planetary Sciences, Kyushu
  University, Fukuoka 819-0395, Japan}

\author[0000-0002-0857-3992]{Shin-ichi Takehiro}
\affiliation{Research Institute for Mathematical Sciences, Kyoto
  University, Kyoto 606-8502, Japan}

\author[0000-0003-4745-4520]{Yoshi-Yuki Hayashi}
\affiliation{Center for Planetary Science, Kobe University, Kobe
  650-0047, Japan}
\affiliation{Department of Planetology, Kobe University, Kobe
  657-8501, Japan}

\begin{abstract}
Direct-imaging techniques of exoplanets have made significant progress
recently, and will eventually enable to monitor photometric and
spectroscopic signals of earth-like habitable planets in the future.
The presence of clouds, however, would remain as one of the most
uncertain components in deciphering such direct-imaged signals of
planets. We attempt to examine how the planetary obliquity produce
different cloud patterns by performing a series of GCM (General
Circulation Model) simulation runs using a set of parameters relevant
for our Earth. Then we use the simulated photometric lightcurves to
compute their frequency modulation due to the planetary spin-orbit
coupling over an entire orbital period, and attempt to see to what
extent one can estimate the obliquity of an Earth-twin.  We find that
it is possible to estimate the obliquity of an Earth-twin within the
uncertainty of several degrees with a dedicated 4 m space telescope at
10 pc away from the system if the stellar flux is completely
blocked. While our conclusion is based on several idealized
assumptions, a frequency modulation of a directly-imaged earth-like
planet offers a unique methodology to determine its obliquity.
\end{abstract}

\section{Introduction \label{sec:intro}}

Direct-imaging of Earth-like planets is a quite challenging but
indispensable technique to revolutionize our understanding of planets
in the near future.  The amplitude modulation of a photometric
lightcurve from a {\it color-changing dot} is sensitive to its surface
pattern, and thus would reveal the presence of lands, oceans, clouds
and even vegetation on the surface of the planets
\citep[e.g.,][]{1993Natur.365..715S, 2001Natur.412..885F,
  2009ApJ...700..915C, 2009ApJ...700.1428O, 2010ApJ...715..866F,
  2011ApJ...738..184F, 2019asbi.book..441S,Rushby2019}.  Indeed,
continuous monitoring of oblique planets over their orbital periods
may even enable one to reconstruct their two-dimensional surface map
\citep{2010ApJ...720.1333K,2011ApJ...739L..62K,
  2012ApJ...755..101F,2018AJ....156..146F}. The feasibility of the
mapping has recently been tested using continuous Earth observations
by Deep Space Climate Observatory orbiting at an altitude of 150 km
\citep{2018AJ....156...26J,2019ApJ...882L...1F,aizawa2020}.

In addition, the lightcurve carries complementary information for the
planet as well. The auto-correlation analysis of the photometric
variation roughly provides us the rotation period of the planet
\citep{2008ApJ...676.1319P}. The obliquity can also be inferred from a
simultaneous fitting of the spin vector and planet surface
\citep[e.g.][]{2010ApJ...720.1333K,2016MNRAS.457..926S,2018AJ....156..146F}.
Such dynamical parameters of the planet are of interest for a general
circulation modeling of Earth-like planets
\citep[e.g.][]{2015ApJ...804...60K,2018AJ....155..266D,2019ApJ...883...46K}.

Strictly speaking, an apparent photometric period observed by a
distant observer is not necessarily identical to the true spin
rotation period due to the planetary orbital motion.  This is related
to the reason why a sidereal day of our Earth $P_{\rm spin}$ is
approximately $365.24/366.24 \times 24 \approx 23.934$ hours, which
corresponds to the true spin frequency $f_{\rm spin} \approx 1.00274$
[day$^{-1}$], instead of the $f_{\rm spin, heliocentric} = 1$
[day$^{-1}$].  The difference between the observed and true spin
rotation frequencies, $f_{\rm obs}$ and $f_{\rm spin}$, is
time-dependent, and sensitive to the geometrical configuration of the
system including the planetary obliquity, $\zeta$, the inclination of
the planetary orbital plane for the observer, $i$, and the observer's
direction (the orbital phase angle $\Theta_{\rm eq}$ measured from the
ascending node, for instance).

Thus the corresponding frequency modulation of the periodicity in the
lightcurve may reveal those parameters, through the presence of
  the large-scale inhomogeneity of the surface.  We emphasize that the
  frequency modulation signal is much less sensitive to the specific
  distribution pattern of the surface than the amplitude
  modulation. \citet[][hereafter K16]{2016ApJ...822..112K} proposed
a novel idea to measure the planetary obliquity from the frequency
modulation, and demonstrated its feasibility successfully using a
static cloud-subtracted Earth model.

\begin{figure}[htbp]
 \centering \includegraphics[width=10cm]{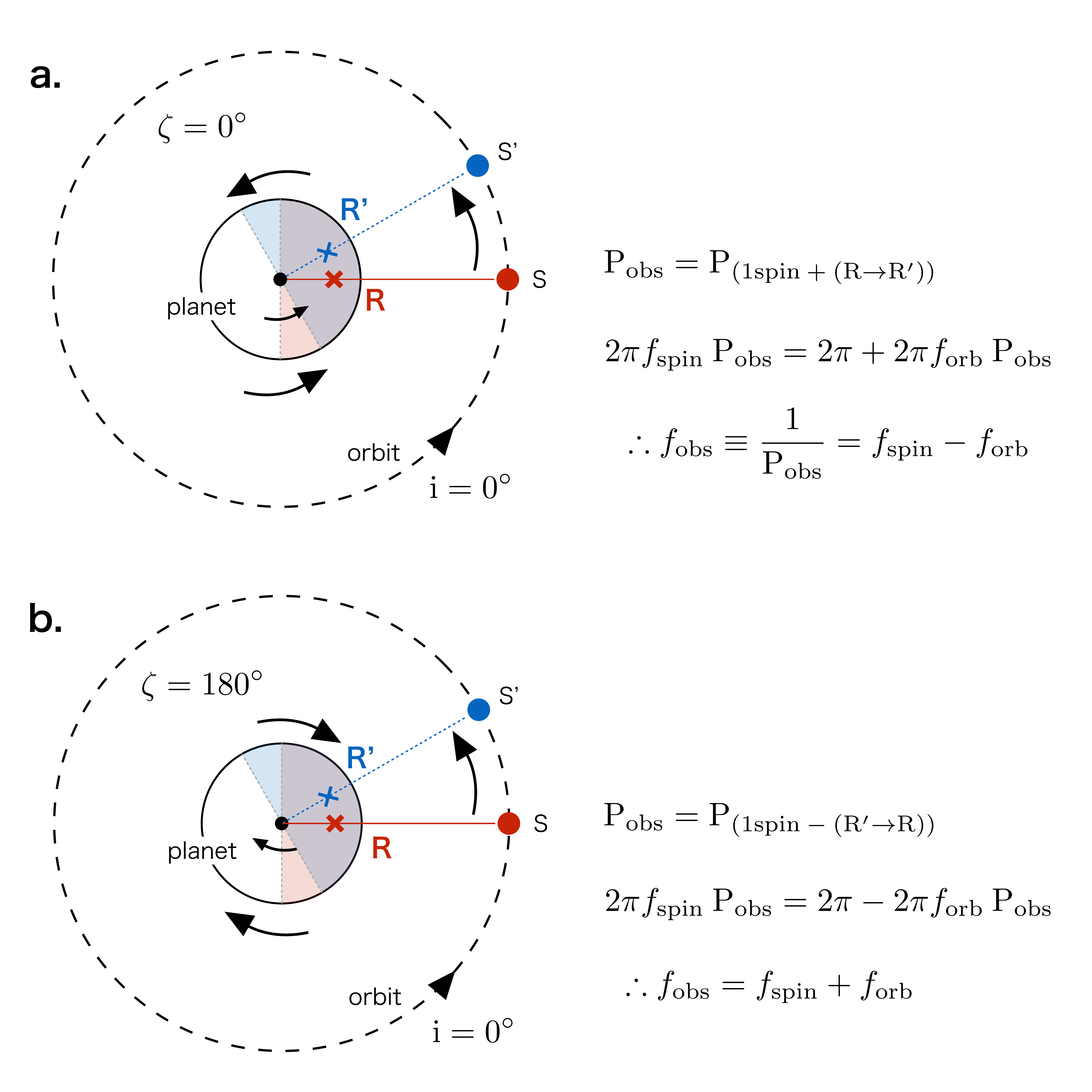}
\caption{A schematic illustration for the periodicity in the
  photometric lightcurve of a planet observed from the direction of
  $i=0^\circ$ relative to the normal vector of the orbital plane.
  Panels a and b indicate prograde ($\zeta=0^\circ$) and
  retrograde ($\zeta=180^\circ$) planets, respectively, in a
      {\it geocentric} frame.  The host star moves around the planet
      from S to S' after one heliocentric day of the planet. The
      illuminated and visible part of the planet from a face-on
      observer (shaded region) changes accordingly, and the reflective
      point at each epoch also moves from R to R'.}
\label{fig:fig1}
\end{figure}

The basic principle of frequency modulation can be understood from
Figure \ref{fig:fig1}.  For a perfectly prograde planet
($\zeta=0^\circ$), the illuminated and visible part of the planet
viewed from a face-on observer ($i=0^\circ$) moves along the same
direction of the planetary spin (Panel a).  The reflective point, at
which the reflected flux of the star is maximal on the planetary
surface, moves accordingly, and thus it takes slightly more than one
spin rotation period $P_{\rm spin}$ for the observer to see the
exactly same part of the planet.  Therefore the observed photometric
variation frequency becomes $f_{\rm obs}=f_{\rm spin}-f_{\rm orb}$.
Applying the same argument, one can easily understand that $f_{\rm
  obs}=f_{\rm spin}+f_{\rm orb}$ for a perfectly retrograde planet
($\zeta=180^\circ$) as illustrated in Panel b of Figure
\ref{fig:fig1}.

\begin{figure}[htbp]
 \centering \includegraphics[width=10cm]{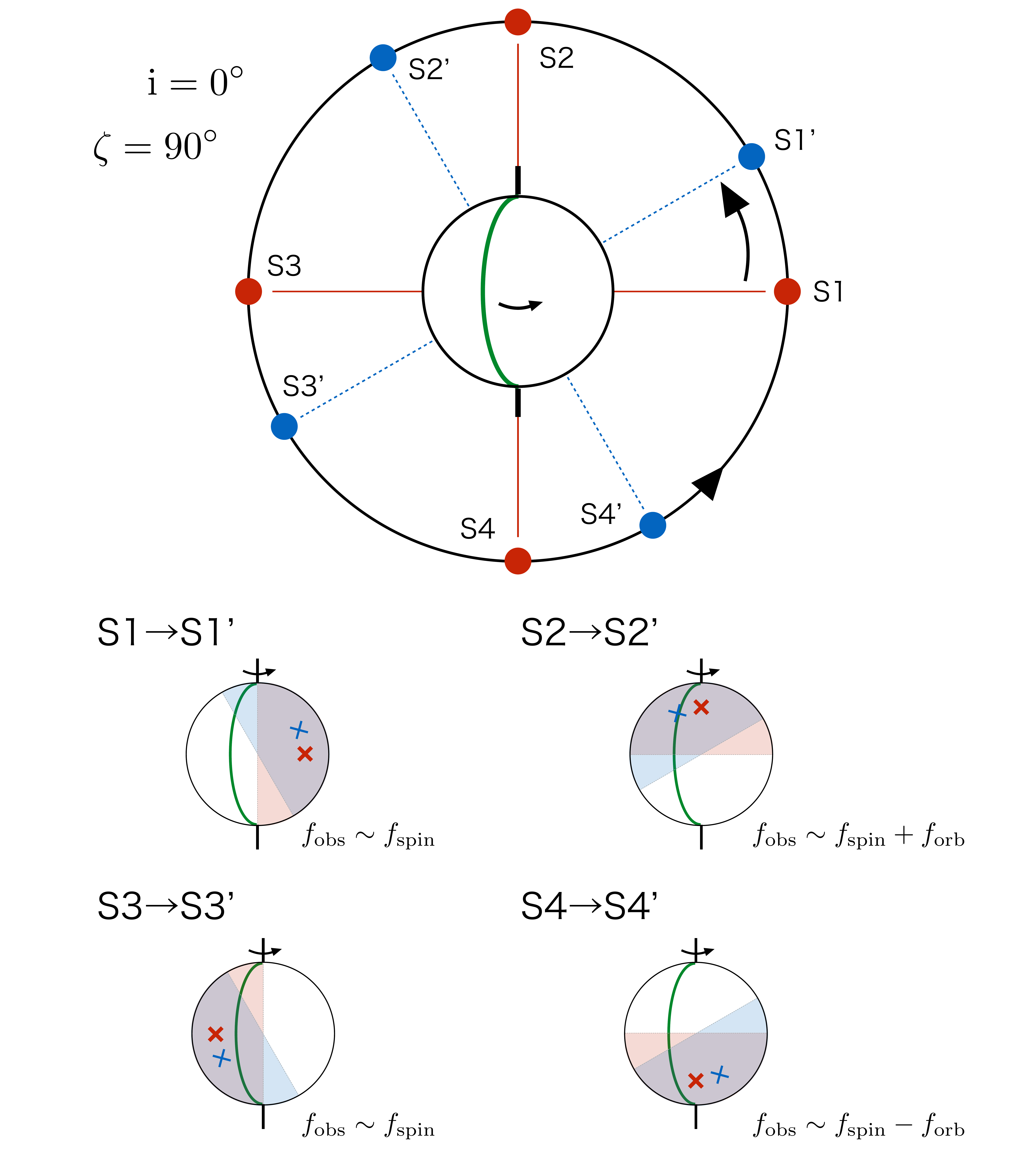}
\caption{Same as Figure \ref{fig:fig1} but for
$\zeta$ = 90$^\circ$ and $i=0^\circ$. } \label{fig:fig2}
\end{figure}

In general, the photometric variation frequency $f_{\rm obs}$ is not
constant and varies according to the mutual geometry between the star
and the planet, leading to a frequency modulation of the photometric
lightcurve of the planet.  Figure \ref{fig:fig2} illustrates an
example of the time-dependent frequency modulation for a
$\zeta=90^\circ$ planet viewed from a distant observer at $i=0^\circ$.
In this case, the motion of the reflective point on the planetary
surface changes the direction relative to the planetary spin axis in a
time-dependent fashion, resulting in the frequency modulation of the
observed period.

When the star is located in S1 (and also in S3), the reflective point
on the planetary surface moves along the constant longitude, and the
planet exhibits a nearly same illuminated and visible part of its
surface after one spin rotation period. This implies that $f_{\rm obs} 
\approx f_{\rm spin}$.  In contrast, when the star is located at
S2 (S4), the reflective point after one spin rotation period moves
slightly westward (eastward), leading to $f_{\rm obs} \approx f_{\rm
spin}+f_{\rm orb}$ ($f_{\rm obs} \approx f_{\rm spin}-f_{\rm orb}$).

While the above frequency modulation is basically determined by the
geometrical configuration of the system characterized by $\zeta$, $i$,
and $\Theta_{\rm eq}$ as mentioned above (see also Figure
\ref{fig:fig3} below), the most important uncertain factor in modeling
the lightcurve is the time-dependent cloud pattern.  A planet
completely covered by the thick homogeneous clouds, for instance, does
not exhibit any photometric variation, and thus one cannot probe the
surface information at all. In the case of our Earth, approximately
50-60 percent of the surface is covered by clouds on average. Thus, it
is not clear to what extent the interpretation from the frequency
modulation of the lightcurve is affected or even biased by the
properties and time-dependent distribution pattern of clouds.

Since the planetary obliquity is supposed to sensitively change the
cloud pattern among others, the feasibility study of the obliquity
measurement from the frequency modulation requires a self-consistent
modeling of clouds over the entire surface of a planet.  This is why
we perform the GCM (General Circulation Model)\footnote{``General
    Climate Model'' is also referred to as GCM. The two terms are
    often used interchangeably, but sometimes ``General Circulation
    Model'' is more specifically implies a part of modules in
    ``General Climate Model''.  In this sense, our model may be
    referred to as ``an atmospheric General Circulation Model'', but
    we do not distinguish between them in the present paper.}
simulation and analyze the simulated lightcurves for different
planetary obliquities.

The rest of the paper is organized as follows.  Section
\ref{sec:method} describes the basic model of the frequency modulation
in the lightcurve, the GCM simulation of the Earth with different
obliquities, and radiation transfer to simulate lightcurves.  Section
\ref{sec:result} shows the analysis method of the frequency modulation
and the result of the frequency modulation signal extracted from
simulated lightcurves.  Finally section \ref{sec:conclusion} is
devoted to the summary and conclusion of the present paper.

\section{Computational Methods \label{sec:method}}
\subsection{Basic strategy to estimate the planetary
  obliquity from photometric variation  \label{subsec:strategy}}

\begin{figure}[htbp]
 \centering \centering\includegraphics[width=12cm]{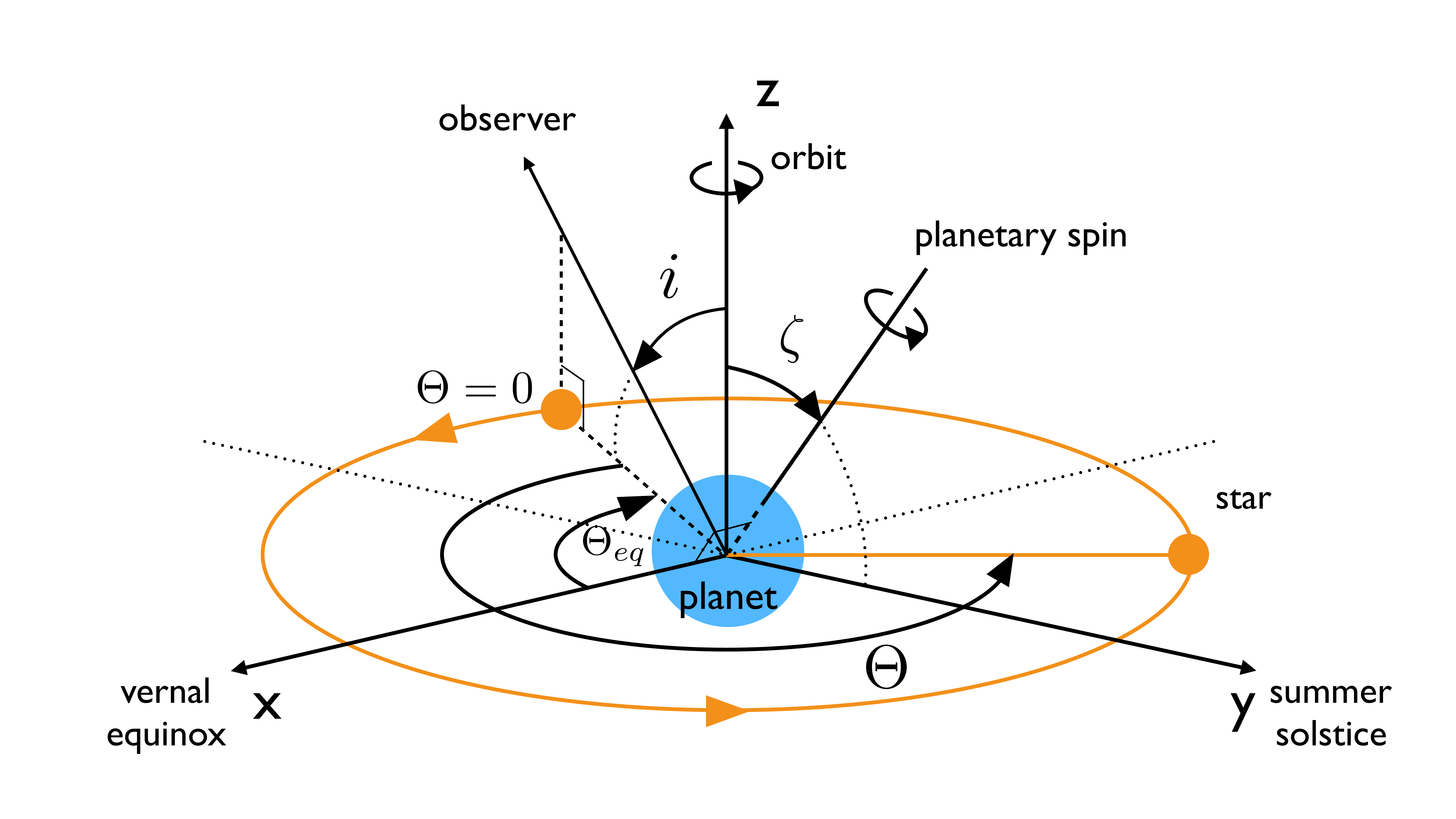}
\caption{A schematic
 configuration of the system in a {\it geocentric} frame.
 The directions of the observer
 and the planetary spin vector do not vary in time,  while the
 direction of the star is time-dependent.} \label{fig:fig3}
\end{figure}

For simplicity, we consider a star-planet system in a circular orbit,
which is schematically illustrated in Figure \ref{fig:fig3}.  In
order to compute the photometric variation of the {\it planet}, it is
convenient to define a {\it geocentric} frame in which the planet is
located at the origin. The stellar orbit defines the $xy$-plane, and
the star orbits around the $z$-axis in a counter-clockwise manner.  The
unit vector of the planetary spin is on the $yz$-plane, and expressed as
$(0, \sin\zeta, \cos\zeta)$ in terms of the planetary obliquity
$\zeta$. Thus the direction of the $x$-axis corresponds to that of the
vernal equinox.

The unit vector toward a distant observer is given by 
$(\cos\Theta_{\rm eq} \sin i, -\sin\Theta_{\rm eq} \sin i, \cos i)$,
where $i$ is the inclination, and $\Theta_{\rm eq}$ is the phase
angle measured clockwise from the $x$-axis ({\it i.e., the vernal
equinox}).

In this frame, the location of the star on the orbit is specified by
its phase angle $\Theta(t)$ measured from the observer's projected
direction.  Since we consider a circular orbit below, $\Theta(t)=2\pi f_{\rm
orb}t$ (mod $2\pi$).

K16 computed the frequency modulation based on a {\it maximum-weighted longitude approximation}, and derived the following formula
for $f_{\rm obs}$ in the case of a circular orbit:
\begin{equation}
\label{eq:fmodel}
f_{\rm model}=f_{\rm spin}+\epsilon_\zeta(\Theta)f_{\rm orb},
\end{equation}
where $\epsilon_\zeta(\Theta)$ is the modulation factor
\footnote{Equation \ref{eq:emodel} is the correct version of equation
  (13) in K16, which contains a couple of typos in signs.},
\begin{eqnarray}
\label{eq:emodel}
\epsilon_\zeta(\Theta)
= \frac{-\cos\zeta\left[1+\cos\Theta\sin i\right]+\sin\zeta\cos i\sin(\Theta-\Theta_{\rm eq})}
{\left[\cos(\Theta-\Theta_{\rm eq})+\sin i\cos\Theta_{\rm eq}\right]^2
+\left[\cos\zeta\sin(\Theta-\Theta_{\rm eq})
-\cos\zeta\sin i\sin\Theta_{\rm eq}-\sin\zeta\cos i\right]^2}.
\end{eqnarray}
We apply the maximum-weighted longitude approximation and derive a
general formula for non-circular orbits in Appendix \ref{sec:epsilon}.
In the present analysis, however, we focus on a circular orbit, and
adopt equation (\ref{eq:emodel}) for the frequency
modulation template.

Following K16, we use the pseudo-Wigner distribution to estimate the
frequency modulation of the photometric variation of a given
lightcurve.  The pseudo-Wigner distribution is the Fourier
transform of the auto-correlation of the data, emphasizing the
periodicity near the time of interest, and reducing the cross terms
and noises. Further detail will be described in Section \ref{sec:result}.

\subsection{GCM simulation of the Earth with different obliquities
  \label{subsec:GCM}}

We would like to emphasize that the main purpose of the present paper
is to examine the feasibility of the planetary obliquity measurement
through the frequency modulation of the lightcurve. The cloud covering
pattern and fraction are important factors that would degrade the
measurement. On the other hand, the precise modeling of the climate is
not supposed to be essential for the feasibility. Therefore various
assumptions and limitations of our current GCM simulation described
below need to be clarified and understood, but do not change the main
conclusion of the present paper.

We use the GCM code {\tt DCPAM5} (the Dennou-Club Planetary
  Atmospheric Model), which has been developed by GFD-Dennou
  Club\footnote{\tt http://www.gfd-dennou.org/, and
      http://www.gfd-dennou.org/library/dcpam/. } for planetary
  climate modeling.  {\tt DCPAM5} has been developed with the aim of
  being able to calculate an atmospheric condition of various
  terrestrial planets, using general formulae as much as possible, by
  excluding properties and modules specific to the Earth 
    \citep[e.g.,][]{2017Icarus..282..1N}.  {\tt DCPAM5} employs the
  primitive equation system assuming that the vertical component of
  the equation of motion is hydrostatic.

\subsubsection{Setup and Sub-grid physical processes}
\label{GCMsetup}

We set the computational grids of $32\times64\times26$ corresponding
to latitudinal, longitudinal, and vertical directions, respectively.
We carry out calculations in the region up to about 6 mbar, which
includes the whole troposphere and a part of the stratosphere.  The
vertical extent of the model domain is enough for our study to express
the generation and motion of clouds because clouds are generated and
advected in the troposphere.  Our simulation resolves the typical
Hadley cell with $\sim 5\time 10$ grids, and thus reproduces the
global meridional circulation observed on the Earth reasonably well.

We use some parameterized physical processes.  In the shortwave
(visible and near infrared, corresponding to the range of incident
stellar flux) radiation process, we take account of absorption by
H$_2$O and CO$_2$, absorption and scattering by clouds, and the
Rayleigh scattering.  In the longwave (mid and far infrared,
corresponding to the range of planetary thermal emission) radiation
process, we take account of absorption by H$_2$O, CO$_2$ molecules and
clouds.  The level-2.5 closure scheme of \citet{1982RvGSP..20..851M}
is used for turbulent diffusion.  The methods of
\citet{1991JApMe..30..327B} and \citet{1995QJRMS.121..255B} are used
for surface flux calculation.  Moist convection is parameterized by
the Relaxed Arakawa-Schubert scheme described in
\citet{1992MWRv..120..978M}.  Large scale condensation (non-convective
condensation) is parameterized by the scheme of
\cite{1991ClDy....5..175L}.  The amount of cloud water is calculated
by integrating a time dependent equation including condensation,
evaporation, advection, turbulent diffusion, and sedimentation of
cloud water.  Extinction rate of cloud water is assumed to be
proportional to the amount of cloud water, and extinction time is
given as an external parameter.  The bucket model of
\cite{1969MWRv...97..739M} is used for soil moisture calculation.  We
use a slab ocean model and its depth to 60 m, the value of
\cite{doi:10.1002/2014JD022659}.

Our simulation is intended to produce a simulated lightcurve for
an Earth-twin but with different obliquity $\zeta$. Thus we
basically adopt the known parameters of the Earth, except for its
obliquity. For simplicity, we set the orbital eccentricity
and the orbital period to  be
$e= 0$ and  $P_{\rm orb}=365.0$ day.

We solve surface temperature and sea ice concentration directly from
our simulation, instead of adopting the observed value for the Earth
with $\zeta$ = 23.44$^\circ$, since those values change with the
different values of $\zeta$.  We use observational data of
surface geological properties,
neglecting that the change of climate also affects those parameters.
  Surface albedo is calculated at each grid point according to the surface
  geological properties, land moisture, and temperature
\footnote{The model codes and related data for the GCM
    experiments are available at
    {\tt http://www.gfd-dennou.org/library/dcpam/sample/}}.

Because our GCM does not include the microphysics of cloud
formation, cloud parameters are fixed to those for the Earth;
effective radius of water and ice cloud particle are set to be
10$\mu$m and 50$\mu$m, respectively.  Lifetime of water and ice
clouds are chosen to be 3240 seconds and 8400 seconds,
respectively.

\subsubsection{Initial Conditions \label{GCMinitial}}

In the present paper, we consider six different values for the
obliquity; $\zeta=0^\circ$, $30^\circ$, $60^\circ$, $90^\circ$,
$150^\circ$, and $180^\circ$. The simulation runs for $\zeta=0^\circ$,
$150^\circ$, and $180^\circ$ start from the isothermal atmosphere of
temperature $T_{\rm init}=280$K and surface pressure $p_{\rm
  s}=10^5$Pa with initially vanishing specific humidity and wind
speed. Then we evolve those three models for 20 years so that they
reach equilibrium. We call this process the relaxation run.

For $\zeta=30^\circ$, $60^\circ$, and $90^\circ$, we first run the
case of $\zeta=15^\circ$ with exactly the same initial conditions
mentioned above for 20 years as the relaxation run.  We adopt the
final result of the relaxation run with $\zeta$ as the initial
condition for the next model with $\zeta+\Delta\zeta$. We set
$\Delta\zeta=5^\circ$.  The system with $\zeta+\Delta\zeta$ becomes
almost in equilibrium after 10 years since the final epoch of the
relaxation run for $\zeta$; the annual mean of the total atmospheric
energy is constant within the level of 0.1\% for the last 5 years.
Thus we stop the relaxation run in 10 years.  We repeat the process up
to $\zeta=90^\circ$. The incremental procedure is mainly to save the
computation time.  In the retrograde runs for $\zeta=150^\circ$ and
$180^\circ$, we skip the incremental procedure, and made sure that the
results reach the equilibrium state after the 20 years relaxation run
directly from the isothermal atmosphere.

The simulated data that we analyze below is computed for an additional
one year after each relaxation run.  We output the physical parameters
every 3 hours for the entire period, which is the required time
resolution for detecting the rotation frequency (corresponding to 24
hours) and its modulation from the simulated lightcurve.

\subsubsection{Climate of earths with different obliquities}\label{GCMresult}

\begin{figure*}
\centering \centering\includegraphics[width=14cm]{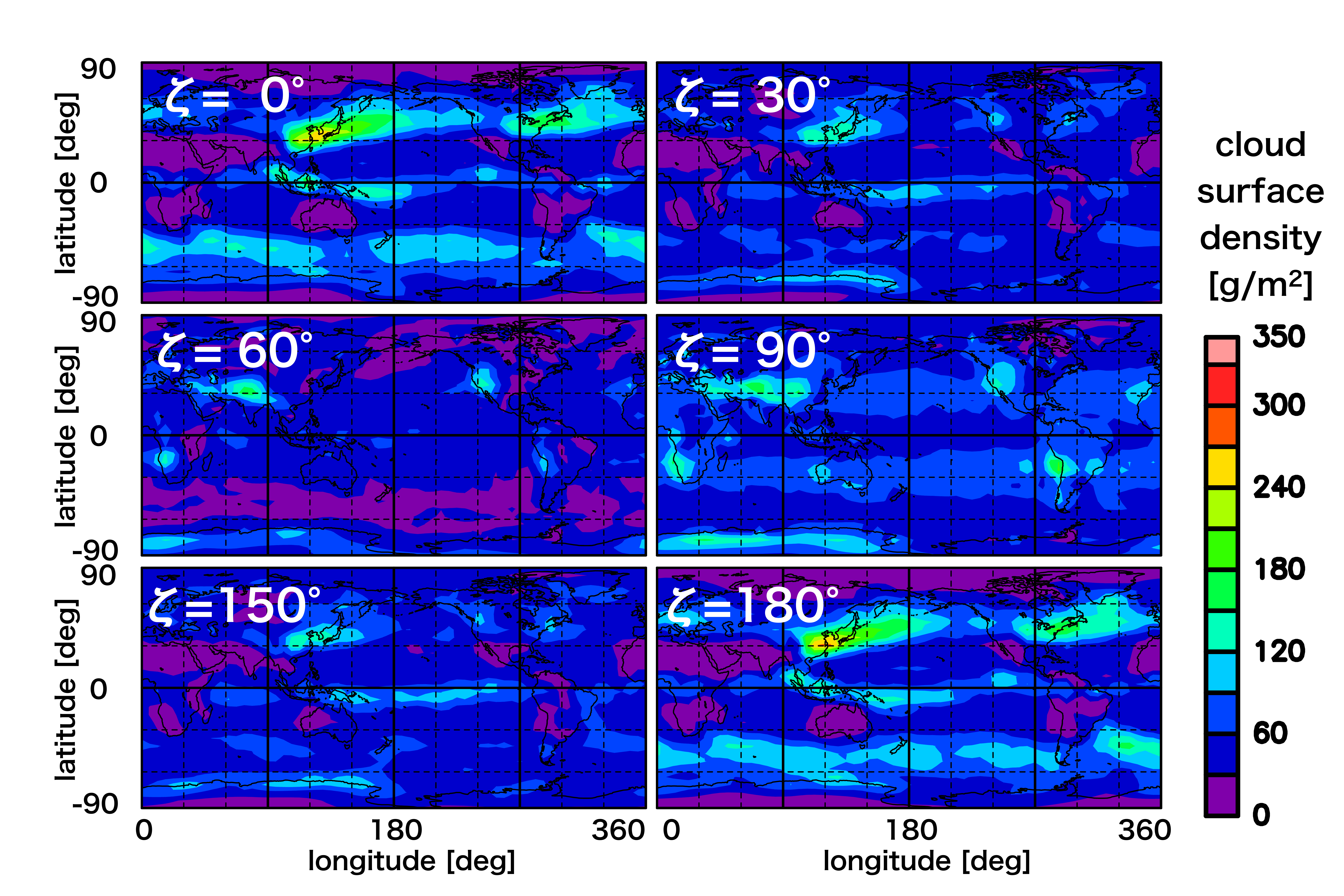}
\caption{Annual mean cloud column density $[{\rm g/m}^2]$
of GCM experiments with six individual obliquities $\zeta$.
\label{fig:fig4}}
\end{figure*}

Figure \ref{fig:fig4} shows the annual mean cloud
column density distribution of planets with different obliquities.
Results for $\zeta\leq$ 30$^\circ$ show the cloud belts on the equator and
mid-latitudes. The clouds around the equator are generated by the
Hadley circulation. This circulation also produces subtropical
highs, which are shown as the partially cloudless continents around
the latitude $\lambda=20-30^\circ$.

The cloud patterns for $\zeta=150^\circ$ and $180^\circ$ are very
similar to those for $\zeta=30^\circ$ and $0^\circ$,
respectively. This is due to the symmetry with respect to the stellar
location for the cases of $\zeta$ and $180^\circ-\zeta$.  The results
for $\zeta=60^\circ$ and $90^\circ$ have the different cloud patterns
due to their atmospheric circulation from the day-side pole to the
equator. The present result is roughly consistent with that shown in
\citet{2003IJAsB...2....1W}, but quantitative comparison is beyond the
scope of this paper.  As we mentioned earlier, however, the precise
modeling of the climate is not the focus of this work.  We plan to
make further comparison elsewhere.

\subsection{Simulated lightcurves \label{subsec:lightcurve}}

\subsubsection{Scattering model and
radiative transfer through the planetary atmosphere}

The total flux of the scattered light from the planet $F(\lambda)$ at
  wavelength $\lambda$ is computed by integrating the intensities $I$
  over the illuminated({\rm I}) and visible({\rm V}) region of the
  planetary surface:
\begin{eqnarray}
\label{eq:RadFlux}
F(\lambda) = \int_{\rm I\cap V}
I(\vartheta_0, \vartheta_1, \varphi; \lambda) \cos\vartheta_1 dS
\frac{1}{D_{\rm obs}^2},
\end{eqnarray}
where $\cos\vartheta_1 dS$ is the projected area element of the
planetary surface viewed by the observer located at a distance of
  $D_{\rm obs}$.

\begin{figure}[htbp]
\centering\includegraphics[width=12cm]{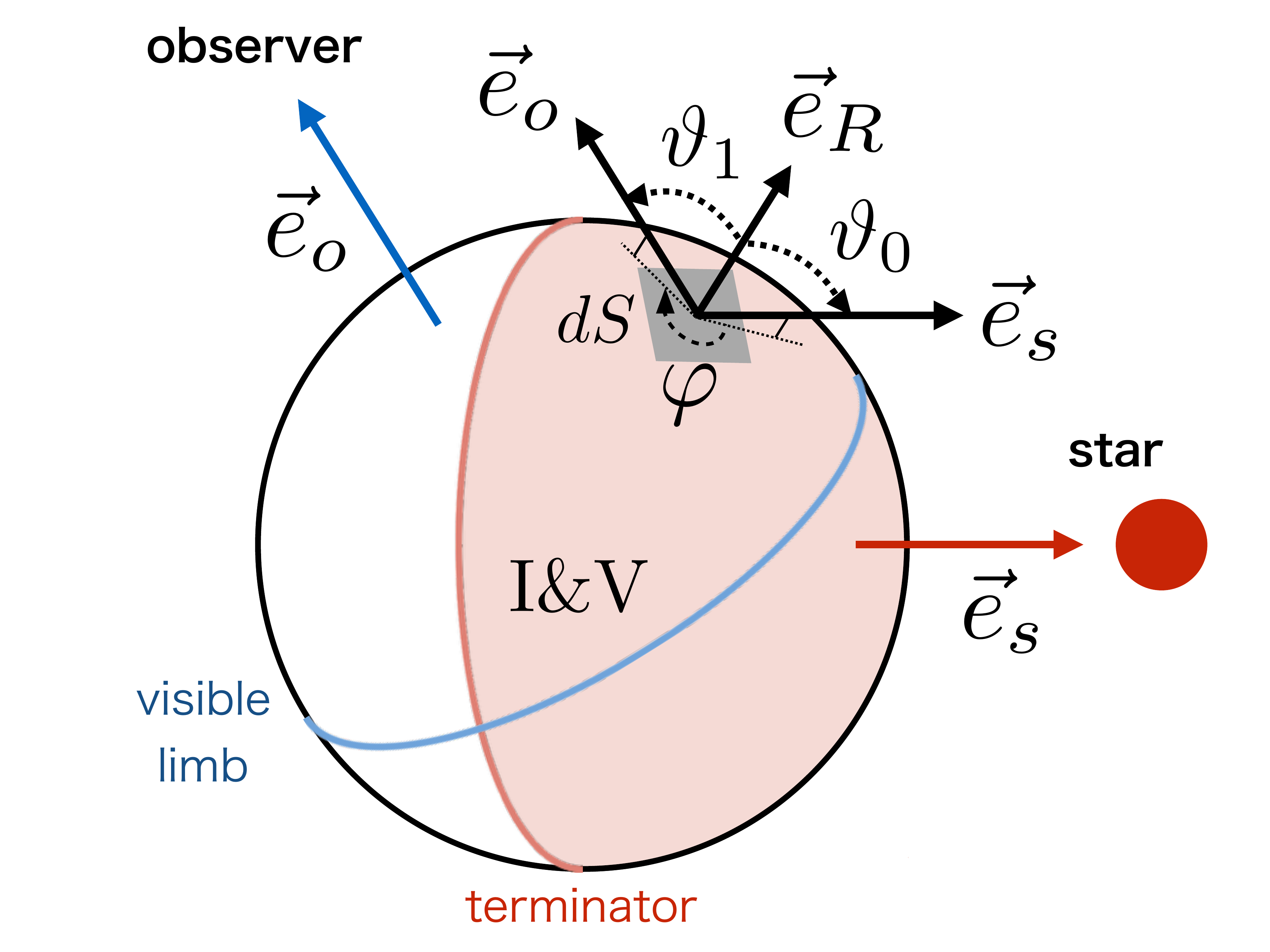}
\caption{A schematic configuration of the scattering geometry.
 \label{fig:fig5}}
\end{figure}

The location of each planetary surface area element is specified
by the three angles ($\vartheta_0$, $\vartheta_1$ and $\varphi$) as
illustrated in Figure \ref{fig:fig5}.
Then the intensity $I$ from the planetary surface area element
is given by
\begin{equation}
\label{eq:intensity}
I(\vartheta_0, \vartheta_1, \varphi; \lambda)
= F_{*,p}(\lambda) \cos\vartheta_0 ~f(\vartheta_0, \vartheta_1, \varphi; \lambda),
\end{equation}
where $F_{*,p}$ is the incident flux and $f$ is the BRDF (bi-directional
reflectance distribution function) that characterizes the scattering
properties of the planetary surface.

Because $f$ includes the entire radiative effects of atmosphere,
clouds and solid/liquid planetary surface, we need to perform a
numerical radiative transfer calculation through the planetary
atmosphere.  For that purpose, we compute $f$ using a public code {\tt
  libRadtran} \citep{gmd-9-1647-2016,acp-5-1855-2005}, which solved
the radiative transfer based on various detailed models of optical
properties of Earth's atmosphere, clouds, aerosols, lands, and ocean
\footnote{We use the {\tt libRadtran} version 2.0.1. 
URL:{\tt http://www.libradtran.org/doku.php}}. 

The {\tt libRadtran} provides several different options for specific
models. We choose the following options.
\begin{enumerate}
    \item We choose REPTRAN \citep{2014JQSRT.148...99G} for optical
      properties of the planetary atmosphere.
    \item We compute optical properties of clouds according to
      \citet{1993JCli....6..728H}. We adopt 10 $\mu$m for the
      effective radius of water cloud particles, as assumed in our GCM
      simulation.
    \item We select the Ross-Li BRDF model \citep{JGRD:JGRD3769} for
      land scattering.  We adopt three Ross-Li parameters that are
      required to provide in {\tt libRadtran} from a remote sensing
      project of the Earth, MODIS \citep[MODerate resolution Imaging
        Spectroradiometer;][]{1989ITGRS..27..145S}.  More
      specifically, we choose their data set ``snow-free gap-filled
      MODIS BRDF Model Parameters''.  In doing so, we employ the data
      in March, neglecting the annual variation.  Also, we sample the
      three parameters at the center of each grid on the
      planetary surface ($32\times64$), instead of averaging over the
      entire grid.  We adopt the above approximation just for
      simplicity. 
    \item Since the above particular data set does not have sufficient
      information for Antarctica, we assume the Lambert scattering and
      employ the ice albedos of (0.948, 0.921, 0.891, 0.837, 0.562,
      0.233), corresponding to the six MODIS bands from 1 to 6
      described below. These values of ice albedo is picked from the
      data ``snow-free gap-filled MODIS BRDF Model Parameters'' at
      (N69$^\circ$.20, W39$^\circ$.35).  This approximation is not
      serious because the ice albedos do not change so much depending
      on the area.
    \item We select the ocean reflection BRDF model of
      \citet{1983JQSRT..29..521N} that is implemented in {\tt
        libRadtran}. We choose 4 m s$^{-1}$ for the wind speed at 10 m
      above the ocean.  Further detail can be found in
      \citet{2010ApJ...715..866F}.
    \item Finally, we solve the radiative transfer equation through
      the atmosphere under a plane parallel approximation.  We choose
      DISORT \citep[DIScrete-Ordinate-method Radiative Transfer
        model;][]{1988ApOpt..27.2502S}.
\end{enumerate}

We use the GCM output of water cloud density, ice cloud density,
temperature, air density, and vapor mixing ratio as the input vertical
profiles of atmosphere and clouds for {\tt libRadtran}.  While our GCM
simulations distinguish between ice cloud and water cloud, we regard
the ice cloud as water cloud in {\tt libRadtran} so as to reduce the
computational cost.  For simplicity, we ignore the radiative transfer
outside the region of GCM simulation ($z\sim$ 0-30 km), including
effects due to the upper atmosphere of the planet, exo-zodiacal dust and
the interstellar medium.

\begin{figure*}
\gridline{
        \fig{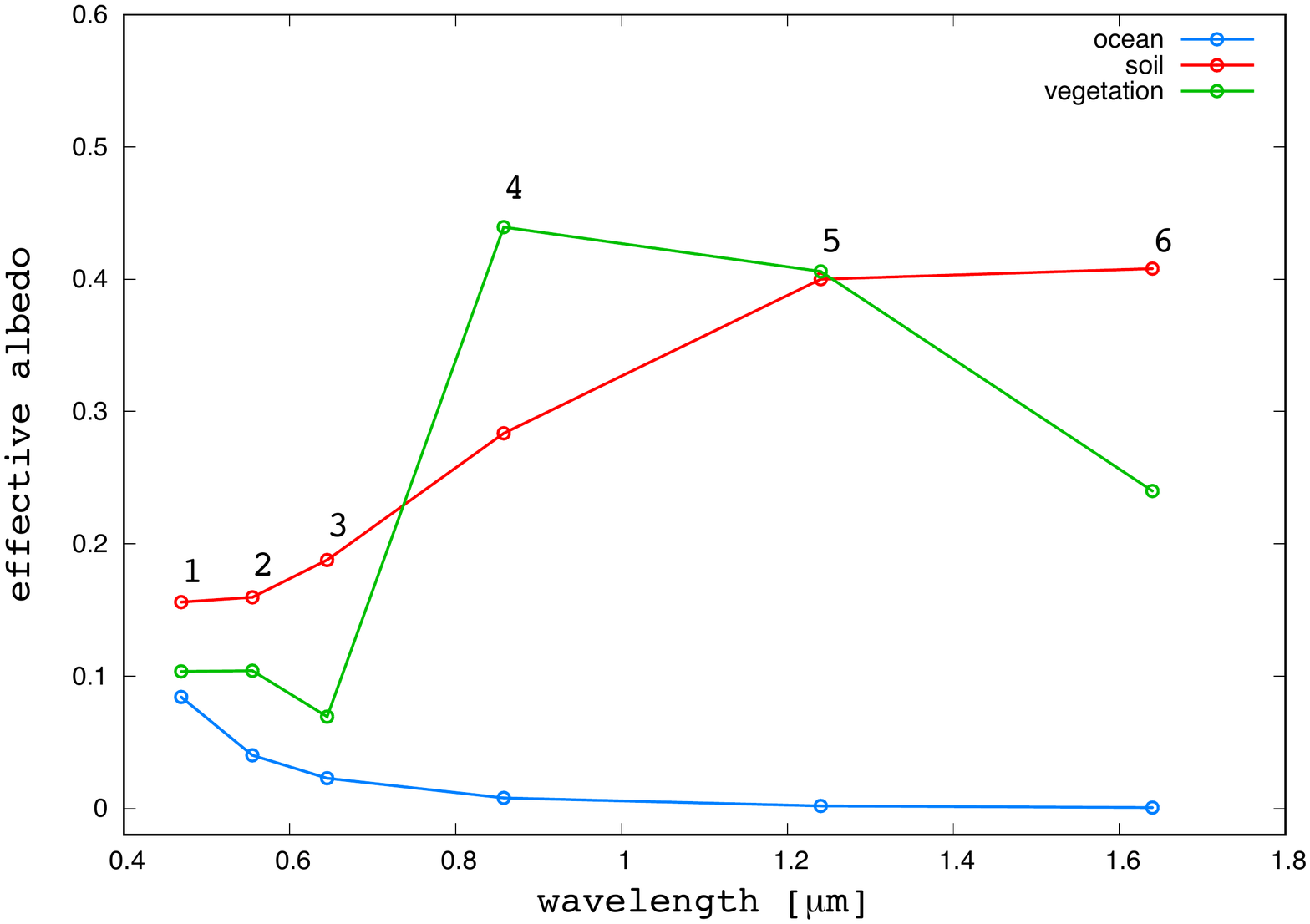}
        {0.5\textwidth}
        {a. Reflectance of the Earth's
          surface components through the Earth's atmosphere.
        }
        \fig{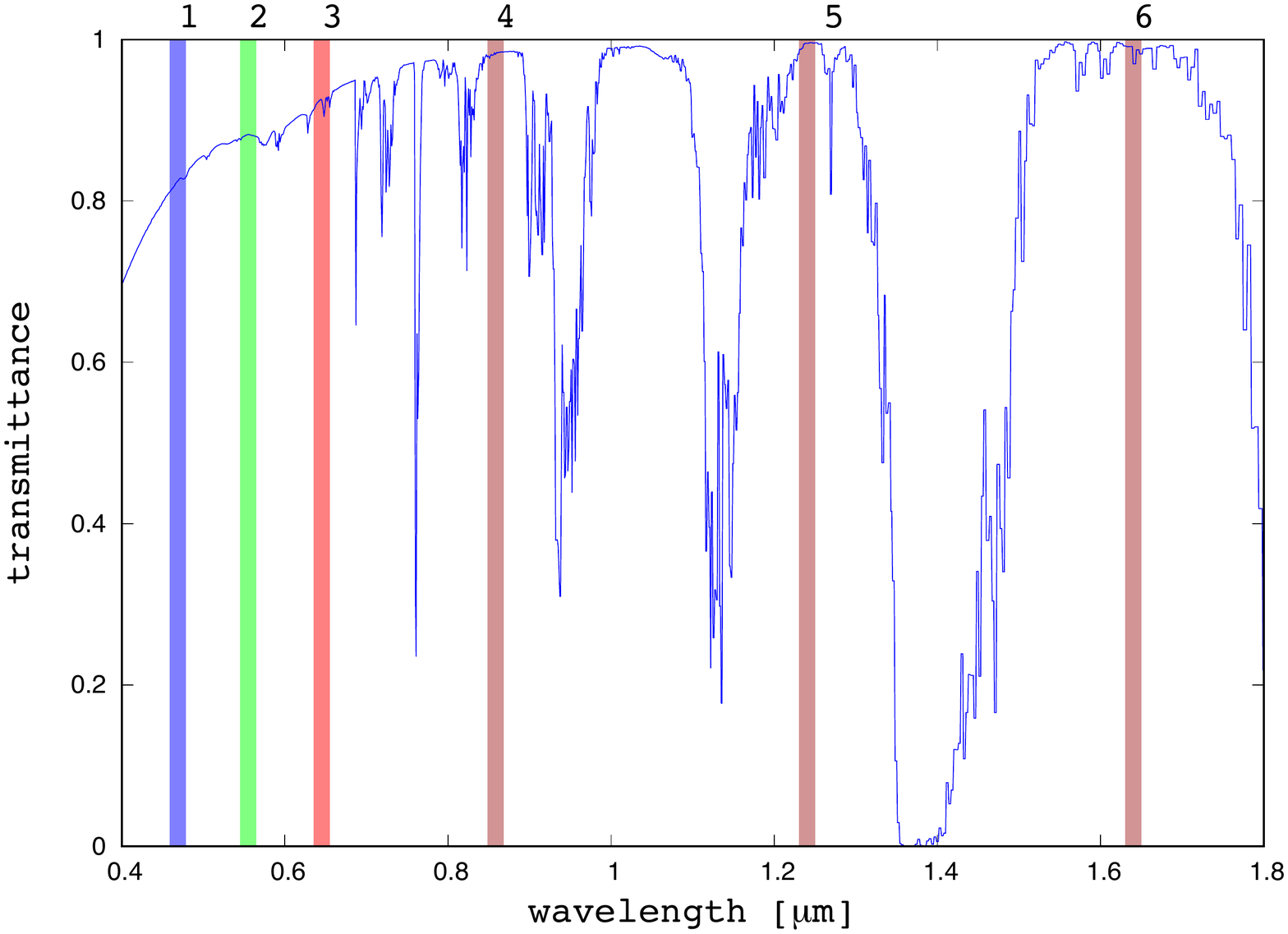}
        {0.5\textwidth}
        {b. Transmittance of the Earth's atmosphere
        and positions of the MODIS photometric bands. 
        }
        }
\caption{Effective albedo and the transmittance of the atmosphere of
  the Earth. Numbers from 1 to 6 above panels indicate numbers of
  bands in the MODIS project (see Table \ref{tab:band}). For
  atmospheric profile, the US Standard Atmosphere
  \citep{1986aacp.book.....A} without cloud is used. \label{fig:fig6}}
\end{figure*}

We compute the intensity in six photometric bands centered at the
wavelengths of the MODIS bands (Table \ref{tab:band}) but with an
expanded bandwidth of $\Delta\lambda=0.1\mu$m.
\begin{table}[htb]
 \caption{Photometric bands of our mock observation}
  \centering
   \begin{tabular}{cc} \hline
band number & MODIS central wavelength\\ \hline
1 &  0.469 $\mu$m \\
2 &  0.555 $\mu$m \\
3 &  0.645 $\mu$m \\
4 &  0.858 $\mu$m \\
5 &  1.240 $\mu$m \\
6 &  1.640 $\mu$m \\ \hline
   \end{tabular}
  \label{tab:band}
\end{table}
The MODIS project selected their photometric bands so as to
characterize the reflection properties of the Earth's surface by
remote sensing. Figure \ref{fig:fig6}a shows examples of effective
albedo (reflectance) spectrum for different components of the Earth's
surface; soil, vegetation, and ocean.  Three bands (1-3) roughly
correspond to the visible color of blue, green, and red, respectively.
Figure \ref{fig:fig6}a exhibits a clear difference among the three
components, ocean, soil, and vegetation.  Incidentally, the MODIS
project chooses 3 near-IR bands that correspond to observational
windows of Earth's atmosphere (Figure \ref{fig:fig6}b).

As we have already emphasized, the cloud distribution is the most
important ingredient in our mock simulation. In order to examine the
dependence on their properties, we generate a simple cloud
distribution as follows. Our GCM result for $\zeta = 30^\circ$
indicates that the simulated cloud distribution has typical column
densities of $0.040^{+0.050}_{-0.025}$ kg/m$^2$.  Thus we
redistribute all the clouds homogeneously within 0.0-0.3, 0.5-1.0, and
3.0-8.0 km, which roughly correspond to the typical heights for mist, lower clouds, and middle clouds for the Earth. 

\begin{figure}[ht!]
\centering\includegraphics[width=10cm]{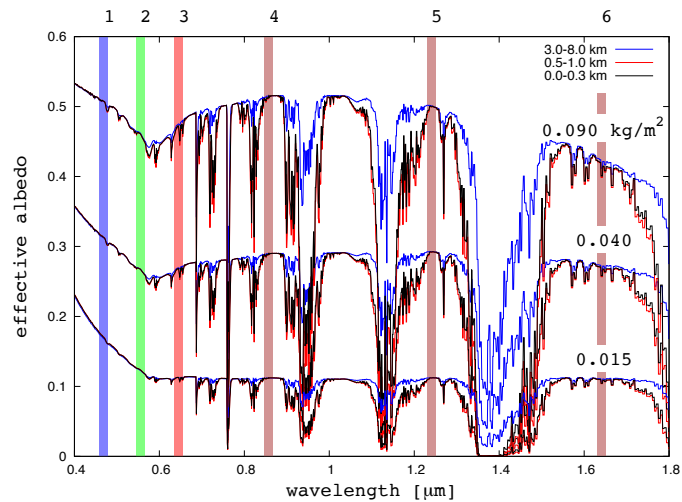}
\caption{Effective albedo of clouds calculated with {\tt libRadtran}
  from our GCM outputs with the obliquity $\zeta=30^\circ$.
\label{fig:fig7}}
\end{figure}

Figure \ref{fig:fig7} shows the resulting effective albedos for
those mock clouds, indicating that the albedos are mainly determined
by the column density, and fairly insensitive to the height of clouds.

\subsubsection{Simulated images and lightcurves of an Earth-twin}

Before performing the frequency modulation analysis, let us present
examples of the apparent images and lightcurves from our mock
observation.

Figures \ref{fig:fig8a} and \ref{fig:fig8b} show the images of an
Earth-twin in January and July, respectively, with different
obliquities viewed from a distant observer at $i=0^\circ$.  Plotted
from left to right are input surface distribution, illuminated and
visible part of the cloudless earth with atmosphere, illuminated and
visible part of the earth with both cloud and atmosphere from our GCM
simulation, and the corresponding cloud distribution.  The arrows
indicate the incident direction of the starlight.

The input surface distribution (the left images) is computed from the
intensity of land alone, neglecting the contribution of the ocean
reflection. The land is assumed to be covered by the US Standard
Atmosphere \citep{1986aacp.book.....A} and land scattering is
approximated by Lambertian. Since those images are just for reference,
we assume the geometric configuration with $(\vartheta_0,
\vartheta_1)=(0^\circ, 0^\circ)$.

The different surface components are illustrated in orange, green
  and blue for continents, vegetation and oceans, respectively. In the
  left images, one may identify North and South America, Eurasia,
  Africa, and Antarctica.  The images of the cloudless earth
relatively well exhibit colors of surface below atmosphere, and also
show an oceanic glint (oceanic mirror reflection) in the illuminating
direction\citep{1993Natur.365..715S,Robinson2010}.  Those
signatures of the surface components are significantly degraded by the
clouds, but one may still identify the presence of the Sahara desert
for $\zeta < 60^\circ$ in Figure \ref{fig:fig8a}, for instance.
Although one may not identify the Sahara desert for $\zeta
  =90^\circ$ in January (Figure \ref{fig:fig8a}), the Sahara desert
  appears in the visible and illuminating part in July (Figure
  \ref{fig:fig8b}). Thus it can be still used a frequency modulation
  indicator partially in a year.

Figures \ref{fig:fig8a} and \ref{fig:fig8b} reconfirm that the cloud
distribution weakens the surface information in photometric
monitoring, but still indicate that the diurnal variation and
possibly its frequency modulation detection are feasible if there
exists a good tracer of the global planetary surface like the Sahara
desert.

\renewcommand{\thefigure}{\arabic{figure}a}
\begin{figure*}[ht!]
	 \centering \includegraphics[width=16cm]{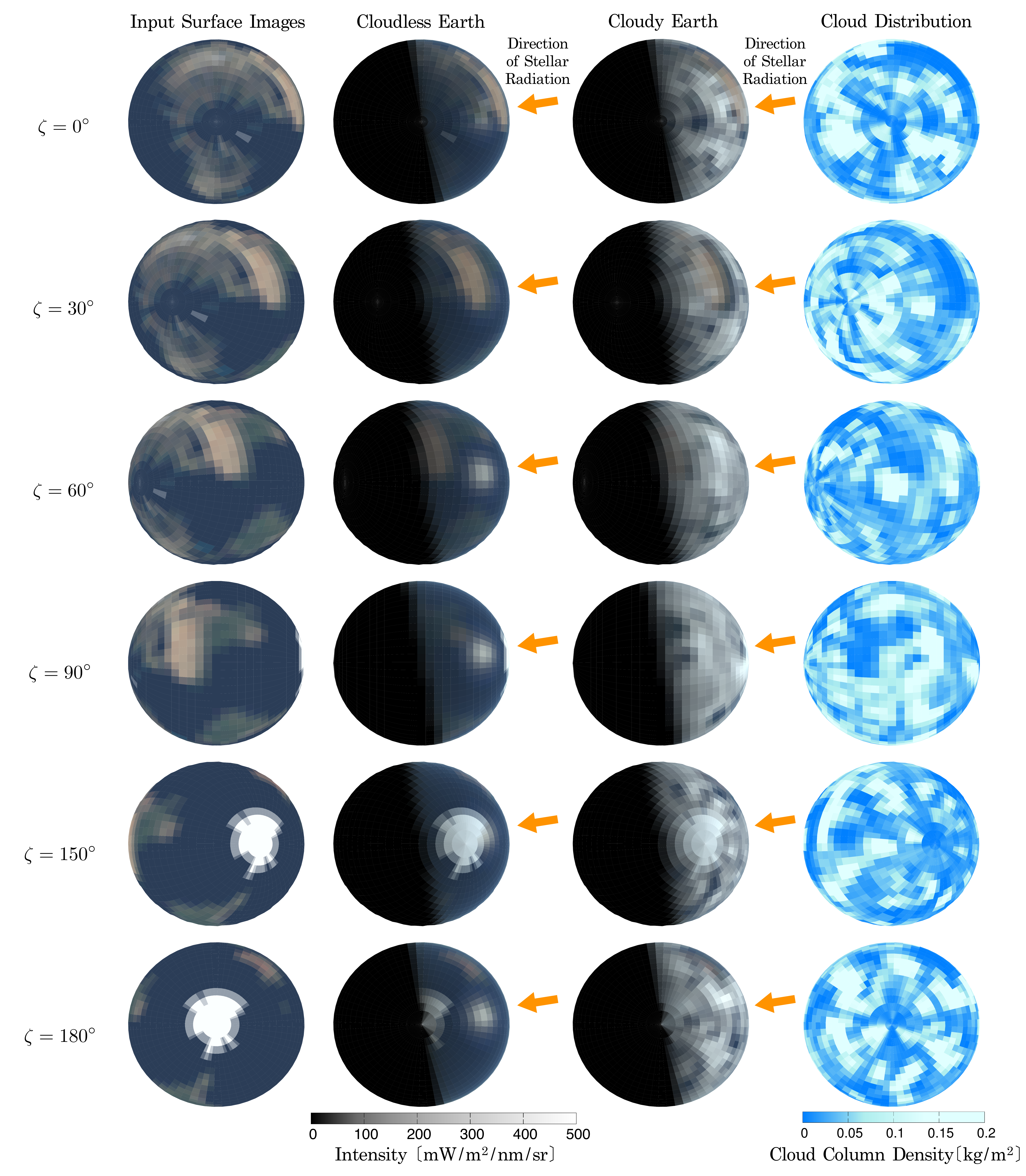}
	 \caption{Images of an Earth-twin from our GCM simulations with
  different obliquities viewed from a distant observer at $i=0^\circ$
  in January.  From left to right, we
  plot the input surface images, illuminated and visible part of the
  cloudless earth with atmosphere, illuminated and visible part of the
  earth with both cloud and atmosphere from our GCM simulation, and
  the corresponding cloud distribution.  The orange arrows show the
  direction of stellar illumination.  We adopt the RGB flux ratio to
  be the intensity ratio of band 3:2:1 (0.645 $\mu$m: 0.555 $\mu$m:
  0.469 $\mu$m) and apply the gamma correction with $\gamma$= 1/2.2 so
  as to roughly represent the apparent colors.}
\label{fig:fig8a}
\end{figure*}

\renewcommand{\thefigure}{\arabic{figure}b}
\addtocounter{figure}{-1}

\begin{figure}
\centering \includegraphics[width=16cm]{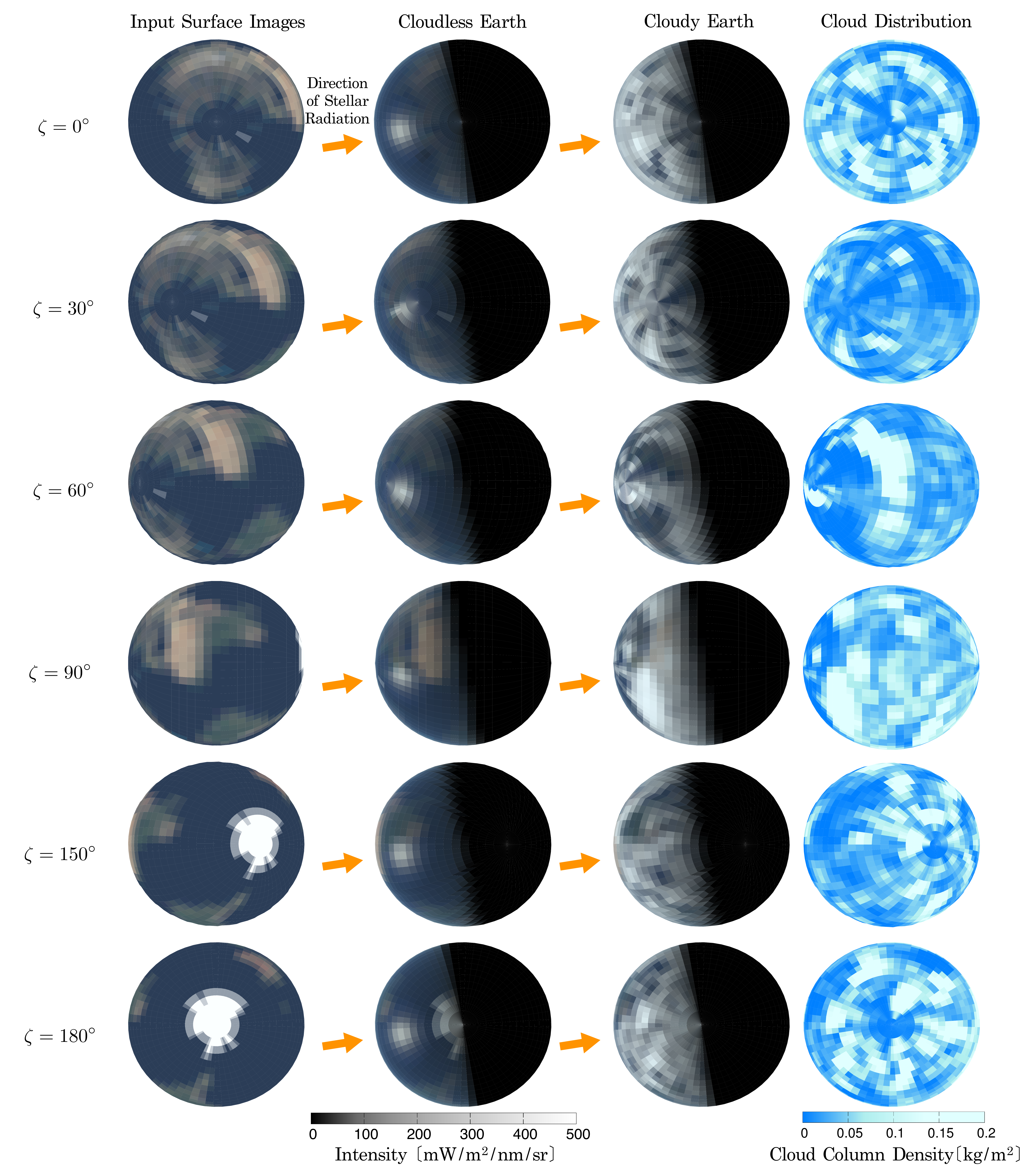}
\caption{Same as Figure \ref{fig:fig8a} but in July.}
\label{fig:fig8b}
\end{figure}

\renewcommand{\thefigure}{\arabic{figure}}

Mock photometric monitoring of the images presented in Figures
\ref{fig:fig8a} and \ref{fig:fig8b} generates the corresponding
simulated lightcurves. Throughout the analysis in what follows, we
consider an observer located at $i=0^\circ$ for simplicity.  Since we
output results of our GCM simulations every three hours, 
we construct simulated lightcurves from those discrete snapshots.
Then we ignore the change of the lightcurve during the three
hours, and construct mock lightcurves sampled every three hours.
While this approximate method significantly affects the lightcurve
variation on a time-scale less than three hours, the variation
around a planetary spin period (24 hours) of our interest is hardly
affected.

\begin{figure*}
\gridline{
        \fig{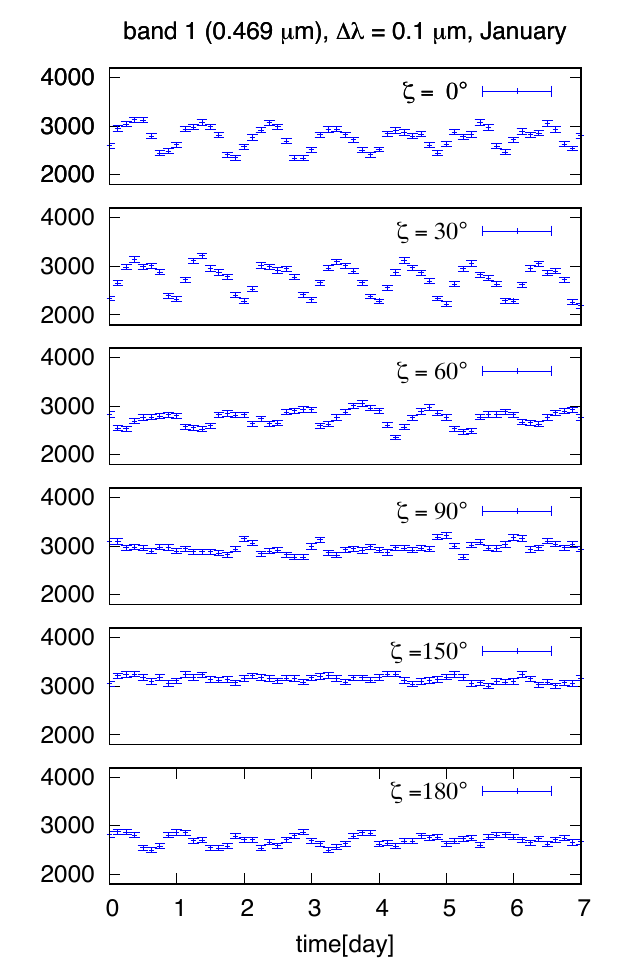}{0.5\textwidth}{}
        \fig{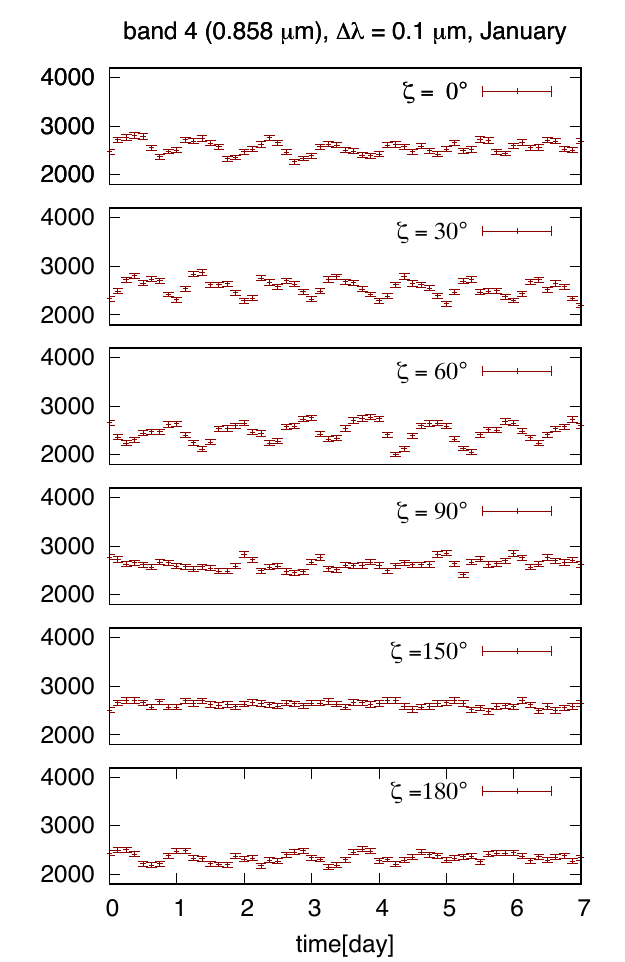}{0.5\textwidth}{}
        }
\caption{Examples of simulated lightcurves
of an Earth-twin with different obliquities for an observer at
$i=0^\circ$. The photon counts $N_{1,0}$ and $N_{4,0}$ correspond to
the set of parameters, $D_{obs}=10\;{\rm pc}$, $D_{\rm tel}=4\;{\rm m}$,
$t_{\rm exp}=3\;{\rm hr}$, and $\Delta\lambda=0.1\,\mu{\rm m}$, and
are scaled as equation (\ref{eq:photon-counts}). The quoted error-bars
consider the photon shot-noise alone.
\label{fig:fig9}}
\end{figure*}

Figure \ref{fig:fig9} shows an example of one-week lightcurves in
January for Earth-twins with different obliquities; left and right
panels correspond to those in band 1 and 4, respectively.  We assume
that the star-planet system is located at a distance $D_{\rm obs}$
away from the telescope of diameter $D_{\rm tel}$ and exposure time of
$t_{\rm exp}$.  In an idealized case where both the light
  from the host star and other instrumental noises are completely
  neglected, the photon counts at band $i$ with the bandwidth of
  $\Delta\lambda$ are scaled as
\begin{equation}
\label{eq:photon-counts}
N_i(t) = N_{i,0}(t)
\left(\frac{D_{\rm obs}}{10\;{\rm pc}}\right)^{-2}
\left(\frac{D_{\rm tel}}{4\;{\rm m}}\right)^2
\left(\frac{t_{\rm exp}}{3\;{\rm hr}}\right)
\left(\frac{\Delta\lambda}{0.1\,\mu{\rm m}}\right) .
\end{equation}
The photon counts in Figure \ref{fig:fig9} correspond to $N_{1,0}$
(left panel) and $N_{4,0}$ (right panel) for the bands 1 and 4 in
equation (\ref{eq:photon-counts}). In practice, we compute
$N_i(t)$ from snapshots every three hours, assuming $t_{\rm exp}=3$hours.

The simulated lightcurves for $\zeta\leq$ 60$^\circ$ exhibit a kind of
diurnal periodicity, which does not reflect the surface information
directly, but comes mainly from the cloud pattern correlated with the
surface distribution.  As $\zeta$ increases ($\zeta \geq 90^\circ$),
the diurnal periodicity is not easy to identify.  As we mentioned in
the above, the Sahara desert played an important role as a tracer of
the planetary rotation, and the annual-averaged cloud pattern is also
correlated to the distribution of the surface components.  This is why
the diurnal periodicity is more visible for photometric monitoring of
the Northern Hemisphere in the case of the Earth.  Although in case of
$\zeta=60^\circ$, the north part of the South America takes the role
as well, it eventually sets out of visible and illuminated region as
$\zeta$ increases.

\section{Time-frequency analysis of simulated lightcurves
  and parameter estimation \label{sec:result}}

Given the simulated lightcurves, we perform the frequency modulation
analysis following K16. In practice, we use a numerical code {\tt
juwvid} to compute the pseudo-Wigner distribution, which is publicly
available from the
web site\footnote{\tt https://github.com/HajimeKawahara/juwvid}.

Our time-frequency analysis proceeds as follows.  First we compute
$N_i(t)$ from our simulated lightcurves every three hours over an
orbital period of one year. Then we sample $N_{i, {\rm obs}}(t)$ from
the Poisson distribution with the expectation value of $N_i(t)$. In
other words, we consider the shot noise alone in the analysis below.
In total, we have $\rm N_{data}=2920$ (=1 year/3hours) data points,
and duplicate the data points with the period of 1 year.

We divide each lightcurve into 73 segments consisting of 40
consecutive data points ({\it i.e.}, 3 hours $\times$ 40 $\approx$5
days). Then we compute the mean $\mu$ and standard deviation $\sigma$
of $N_i(t)$ in each segment, and convert to the {\it normalized}
lightcurve $s(t) \equiv (N_i(t)-\mu)/\sigma$. Finally we compute the
pseudo-Wigner distribution:
\begin{equation}
\label{eq:pseudow}
g(f,t)=\int_{-\infty}^{\infty}
H(\tau)z(t+\tau/2)z^*(t-\tau/2)e^{-2\pi if\tau}d\tau,
\end{equation}
where 
\begin{equation}
\label{eq:analytics}
z(t) = \frac{1}{\pi}\int_0^\infty \tilde{s}(w) e^{iwt}dw
\end{equation}
is the analytic signal of $s(t)$ with $\tilde{s}(w)$ being the Fourier
transform of the normalized lightcurve $s(t)$ in the present case.  We
choose the window function $H(\tau)$ as the following Hamming window
function:
\begin{eqnarray}
\label{eq:Hamming}
h(\tau;T_{\rm w}) =
\left\{
\begin{array}{lr}
0.54 + 0.46\cos(2\pi \tau/T_{\rm w})
&  ~~~~~~ {\rm for} ~~|\tau| \leq T_{\rm w}/2 \\
0 &  ~~~~ {\rm otherwise} .
\end{array}
\right.
\end{eqnarray}
In practice, we adopt $T_{\rm w}=0.25$ year for the
the window width of the Hamming window function.

The pseudo-Wigner distribution is an appropriate time-frequency
distribution for extracting the instantaneous frequency
\citep[e.g.][]{BA25293241}, as explained below. Let us consider a
single mode signal $z = A(t) e^{i \psi(t)}$ with an instantaneous
phase $\psi(t)$, where $A(t) \in \mathbb{R}$ is the amplitude of the
mode. The ideal time-frequency representation is a delta function
$\rho(f,t) = A(t)^2 \delta_D (f - f_{\rm ins}(t)) $, where $f_{\rm
  ins} (t) $ is the instantaneous frequency defined by
\begin{equation}
\label{eq:if}
f_{\rm ins}(t) \equiv \frac{1}{2 \pi} \frac{d \psi(t)}{d t}.
\end{equation}
Then, the inverse Fourier transform of $\rho$ can be written as 
\begin{equation}
\label{eq:invf}
\hat{\rho} (\tau, t) = A(t)^2 e^{2 \pi i f_{\rm ins}(t) \tau}
= A(t)^2 e^{i \tau \psi^\prime(t)}
\approx A(t)^2 e^{i \psi(t+\tau/2) - i \psi(t-\tau/2)}
= z(t+\tau/2) z^\ast(t-\tau/2),
\end{equation} 
where we use the linear approximation $\psi^\prime (t) \approx
[\psi(t+\tau/2) - \psi(t-\tau/2)]/\tau$ in the last two
terms.

Performing the Fourier transform of equation (\ref{eq:invf}) with the
time window, we obtain the pseudo-Wigner distribution. Because the
linear approximation is valid only for the linear frequency modulation
such as $f_{\rm ins}(t) \approx a t + b $ ($a, b$ are constant
values), the width of the window should be chosen to be comparable to
the scale of the non-linear feature of the frequency modulation. The
derivative of equation (\ref{eq:invf}) by $\tau$ at $\tau=0$ provides
\begin{equation}
\label{eq:der}
\frac{d}{d \tau} [z(t+\tau/2) z^\ast(t-\tau/2)] |_{\tau=0}
= i A(t)^2 \psi^\prime(t) = 2 \pi i \int_{-\infty}^{\infty} f \rho(f,t) df.
\end{equation}
Also, the mode amplitude is rewritten as  
\begin{equation}
\label{eq:derampl}
|z(t)|^2 = A(t)^2 = \int_{-\infty}^{\infty} \rho(f,t) df.
\end{equation}
Then, the instantaneous frequency is formally estimated by the
weighted form as,
\begin{equation}
\label{eq:wei}
f_{\rm ins}(t) = \frac{1}{2 \pi} \frac{d \psi(t)}{d t}
= \frac{ \int_{-\infty}^{\infty} f \rho(f,t) df}
{ \int_{-\infty}^{\infty} \rho(f,t) df}.
\end{equation}

In practice, one can estimate the peak value of the pseudo-Wigner
distribution as an instantaneous frequency to avoid the effect of
noise. In this expression, we need a complex-valued signal with
non-negative frequency component of the signal. That is the reason why
we convert a real-valued signal $s(t)$ to the analytic signal $z(t)$
in equation (\ref{eq:analytics}).

We calculate the pseudo-Wigner distribution $g(f,t)$ over the range of
$f_{\rm min}<f<f_{\rm max}$ using equation
(\ref{eq:pseudow}). Specifically we choose $f_{\rm min}=0.98$
[day$^{-1}$] and $f_{\rm max}=1.02$[day$^{-1}$] throughout the
analysis.  Since our lightcurves are sampled every 3 hours, the
corresponding frequency resolution is not good enough to determine the
value of $f_{\rm spin}$ precisely.  Therefore we adopt a non-uniform
FFT scheme \citep{doi:10.1137/S003614450343200X} following K16, and
achieved the frequency resolution of $\delta f= (f_{\rm max}-f_{\rm
  min})/{\rm N}_f$ after applying an appropriate smoothing of the
lightcurves. We choose ${\rm N}_f = 1024$ in what follows, and
  the resulting resolution $\delta f \approx 4\times
  10^{-5}$[day$^{-1}$] is better than the modulation amplitude
  detected in Figure \ref{fig:fig16} by a factor of 100.

\subsection{Single-band analysis \label{subsec:singleband}}

\begin{figure*}
 \centering \includegraphics[width=16cm] {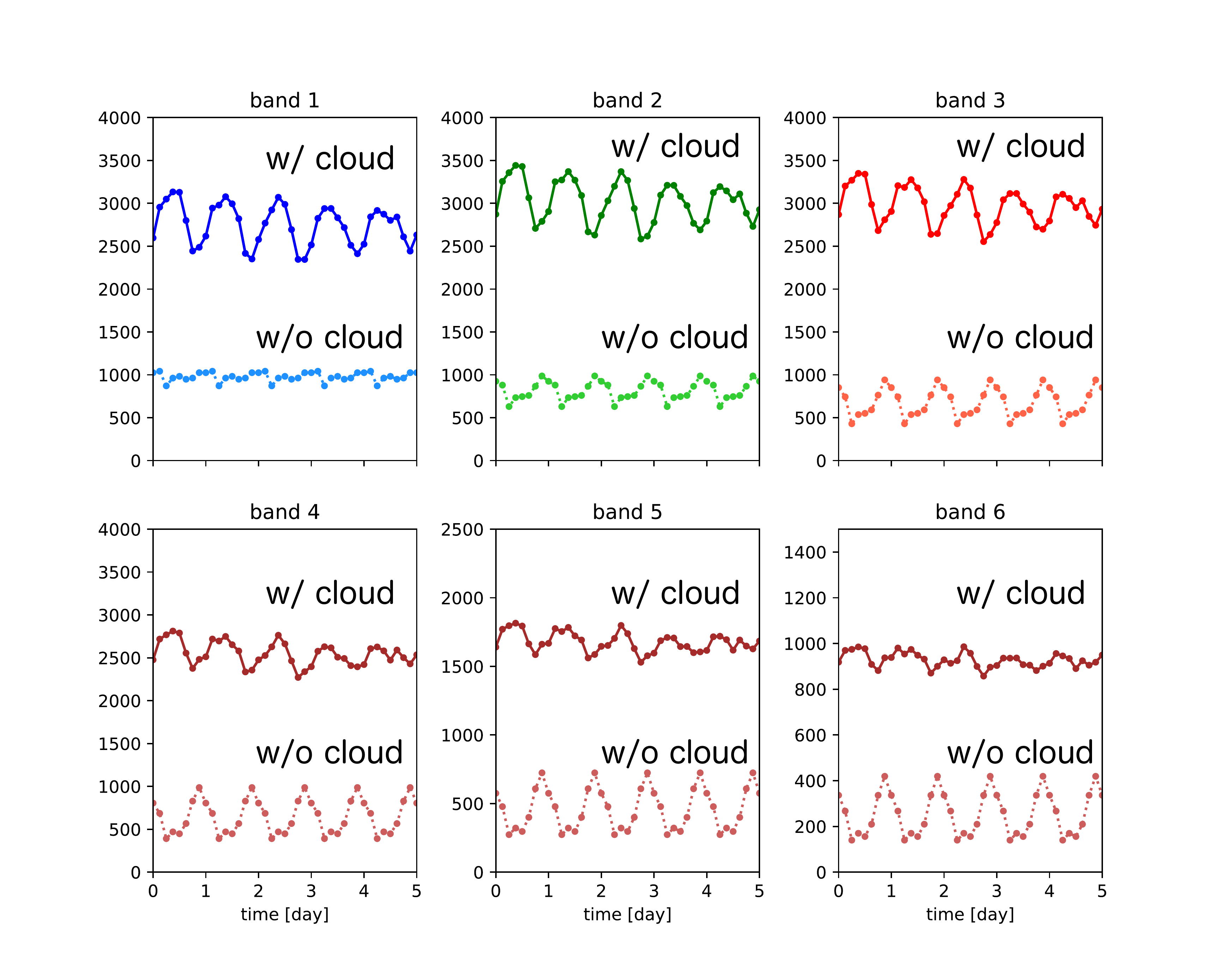}
\caption{Same as Figure \ref{fig:fig9} for $\zeta=0^\circ$, but
in photometric bands 1 to 6 without photometric noise.  Solid and dotted
symbols indicate the lightcurves with and without clouds,
respectively.
\label{fig:fig10}}
 \centering \includegraphics[width=16cm]{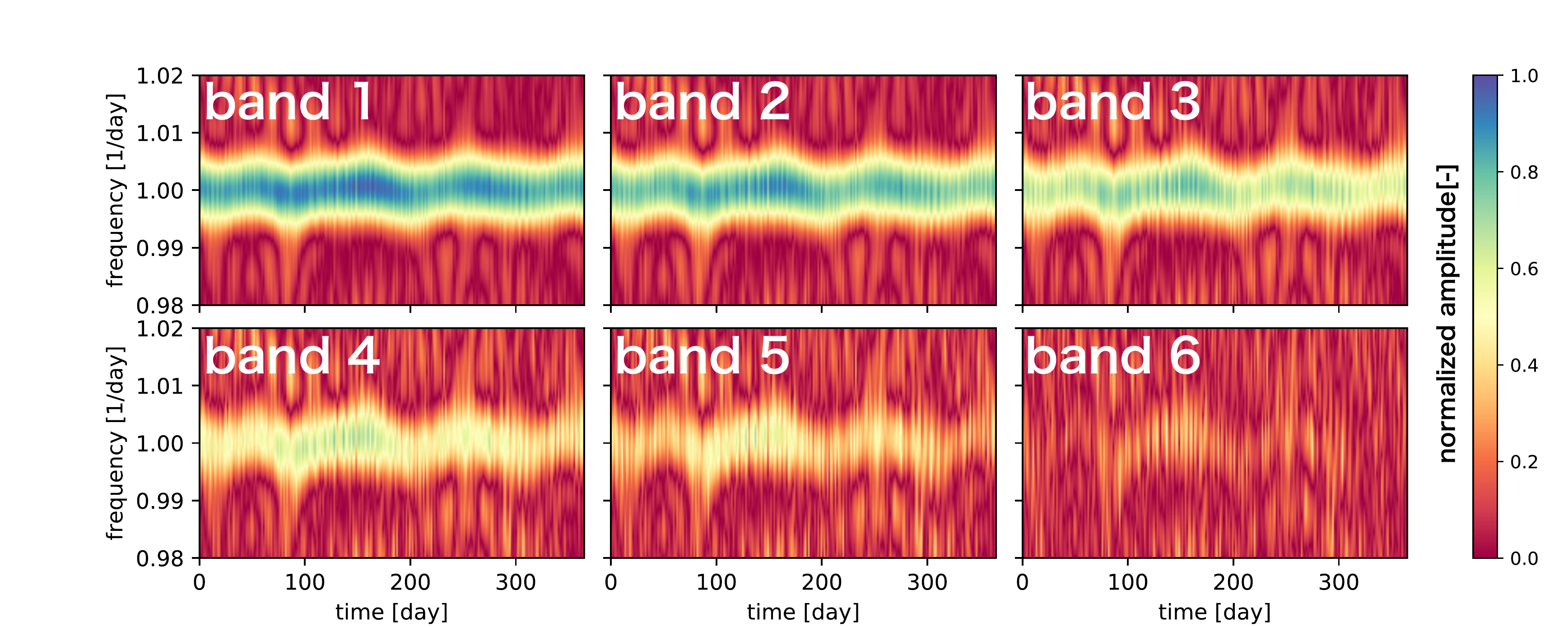}
\caption{Time-frequency representation corresponding to the
noiseless lightcurves of Figure \ref{fig:fig10}.
  \label{fig:fig11}}
\end{figure*}

Consider first the frequency modulation for single-band lightcurves.
Figure \ref{fig:fig10} is similar to Figure
\ref{fig:fig9}, but plots simulated {\it noiseless} (without shot
noise) lightcurves in the photometric bands 1 to 6 for $(\zeta,
i)=(0^\circ, 0^\circ)$.

As clearly indicated by Figure \ref{fig:fig10}, the apparent diurnal
variation in each band originates from the cloud pattern that is
correlated with the land-ocean distribution.  These surface-correlated
clouds were also found in Earth observations by Deep Space Climate
Observatory as the second component of the Principal Component
Analysis \citep{2019ApJ...882L...1F}. While our analysis did not
directly identify the component, it appears to be imprinted in the
diurnal variation in a single band. The amplitude of the single-band
lightcurves is basically determined by the cloud albedo (Figure
\ref{fig:fig7}) multiplied by the incident solar flux. This is why the
amplitude of the diurnal variation with cloud in Figure
\ref{fig:fig10} is relatively large around the visible wavelength
(bands 2 and 3), and declines sharply in the near-infrared (bands 4 to
6).

We note that Figure \ref{fig:fig10} also indicates
the anti-correlation of the lightcurve modulation between cloudless
and cloudy cases. For a cloudless case, the photometric variation is
mainly due to the land component that has larger albedos (Figures
\ref{fig:fig6}a and Figure \ref{fig:fig7}).  Since clouds
are much brighter, however, the photometric variation of a cloudy
case is sensitive to the location of clouds, which tend to avoid the
continent, in particular desert regions, and rather form
preferentially above the ocean. Thus the locations of lands and
clouds are anti-correlated, leading to the anti-correlation
illustrated in Figure \ref{fig:fig10}. This also
explains that the periodic signature of the lightcurve for a cloudy
case is weaker for redder bands, because lands become brighter in
redder bands and compensate the variation due to clouds.

The corresponding color-map for the pseudo-Wigner distribution on the
time-frequency plane (Figure \ref{fig:fig11}) clearly illustrates
the above trend that redder bands have weaker signal.  The color
indicates the absolute value of the time-frequency distribution
density $g(f,t)$, whose maximum value is normalized as unity.  Since
Figure \ref{fig:fig10} is for $\zeta=0^\circ$, the
period for the apparent diurnal variation should be constant, and does
not show any frequency modulation. The tiny frequency modulation $\sim
0.001$ day$^{-1}$ visible in Figure \ref{fig:fig11} is simply due to
the time-dependent inhomogeneous distribution of clouds.

\begin{figure*}
\centering \includegraphics[width=16cm]{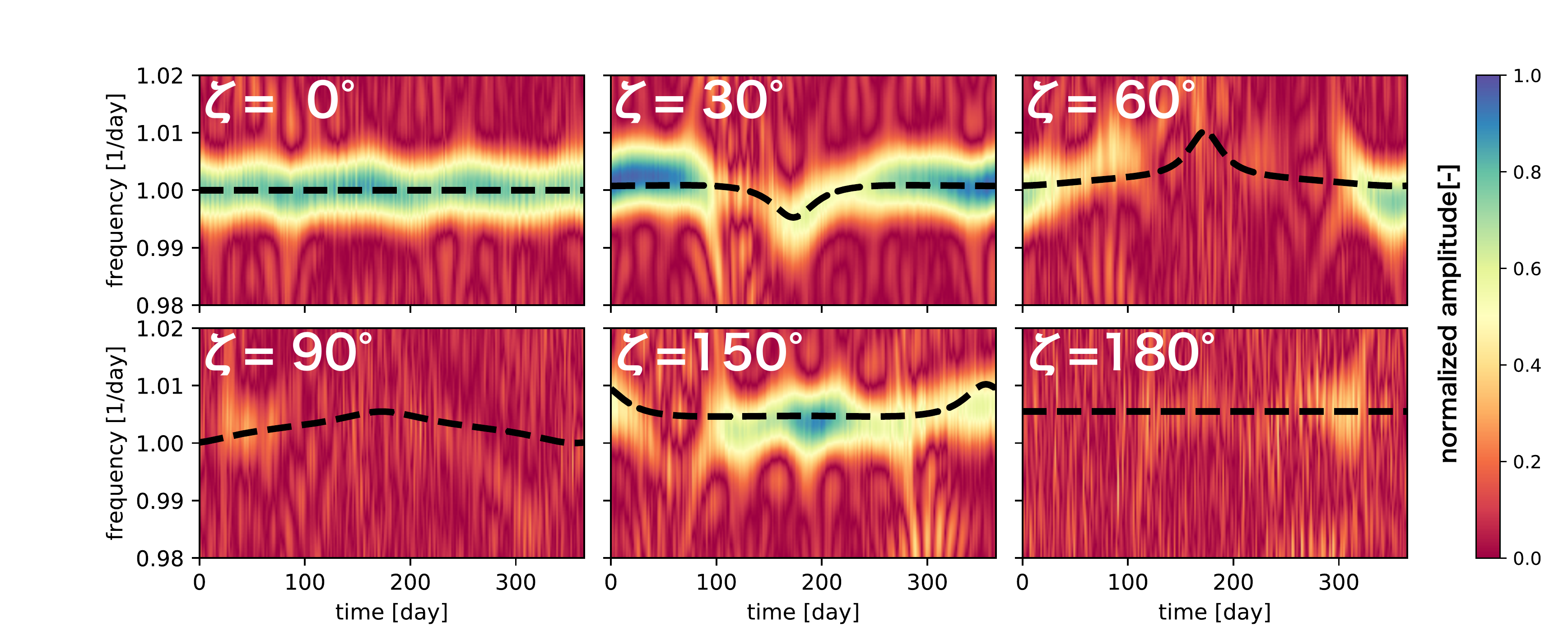}
\centering\includegraphics[width=16cm]{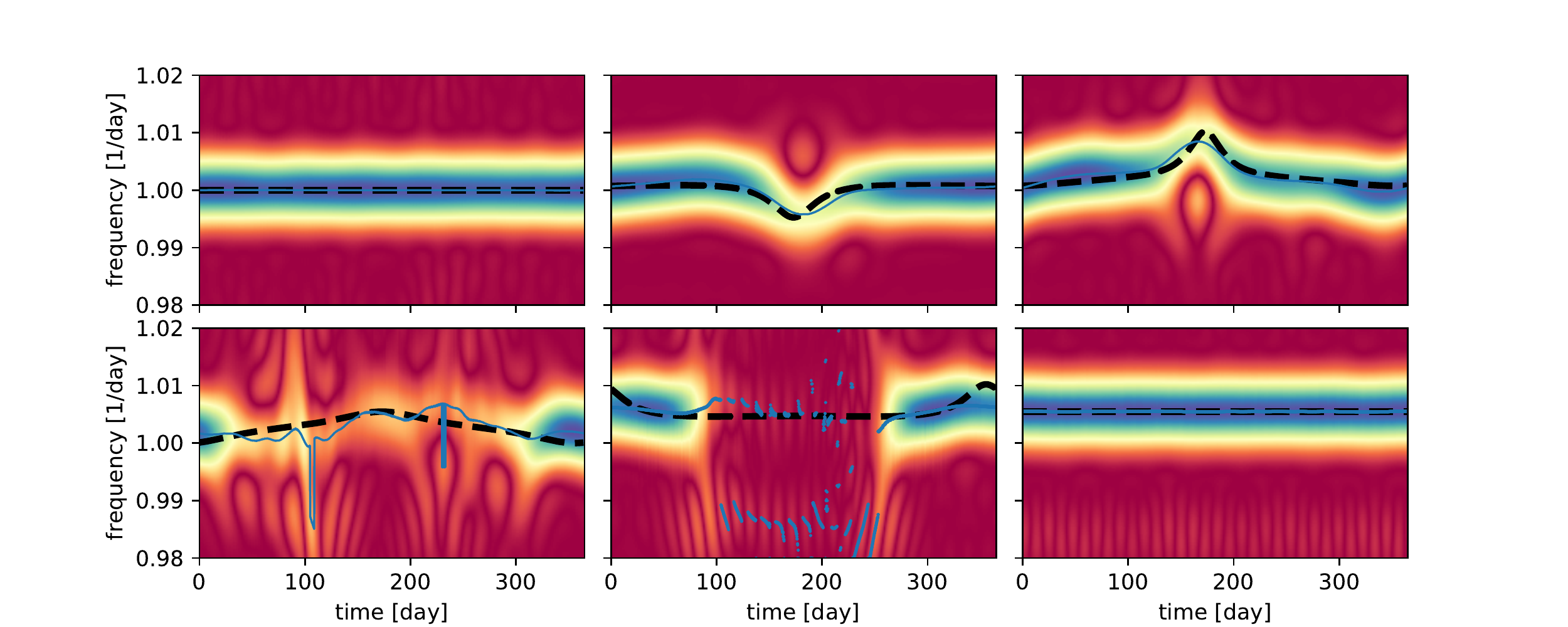}
\caption{Time-frequency representation of band 1 noiseless signal for
  different obliquities ($\zeta=0^\circ$, $30^\circ$, $60^\circ$,
  $90^\circ$, $150^\circ$ and $180^\circ$), corresponding to the
  results with cloud ({\it upper panel}) and without cloud ({\it lower
    panel}) plotted in Figure \ref{fig:fig11}.  Thick dashed lines
  are the prediction based on the maximum-weighted longitude
  approximation model, $f_{\rm model}(t)$; see equations
  (\ref{eq:fmodel}) and (\ref{eq:emodel}) in the main text.
  Thin blue points indicate $f_{\rm data, max}(t)$, the frequency
  corresponding to the maximum value of $g(f,t)$ over a range of
  $f_{\rm min}<f<f_{\rm max}$ at each epoch; see equation
  (\ref{eq:R1}).
\label{fig:fig12}}
\end{figure*}

Consider next the time-frequency representation of the band-1
lightcurves for different obliquities (Figure \ref{fig:fig12}).  We
adopt band 1 because it produces the clearest ridge on time-frequency
representation in Figure \ref{fig:fig11}.  The dashed lines show the
model frequency modulation $f_{\rm model}(t)$, equation
(\ref{eq:fmodel}).  The signature of the frequency modulation from the
single-band lightcurves is not strong, and barely identifiable only
for $\zeta \leq 30^\circ$. Though the amplitude of frequency
modulation is zero for $\zeta = 0^\circ$, the signature of the {\it
  constant} apparent frequency is visible clearly.  This obliquity
dependence reflects the specific distribution pattern of land and ocean
on the Earth. As shown in the $\zeta=150^\circ$ image in Figure
\ref{fig:fig8a}, the illuminated and visible part in winter is
dominated by Antarctica, and there is no significant diurnal variation
in the lightcurve. On the other hand, in summer, Antarctica is almost
invisible and parts of Africa and South America generate the diurnal
variation signal instead.

\subsection{Multi-band analysis \label{subsec:multiband}}

As shown in Section \ref{subsec:singleband}, single-band analysis does
not properly extract the information of the correct frequency
modulation due to the anti-correlation between lands and clouds.  In
order to detect the diurnal period due to the planetary surface
distribution, therefore, we need to remove the time-dependent cloud
pattern as much as possible.

As inferred from the wavelength dependence of albedos for lands
and clouds, bands 1 and 4 are mainly sensitive to clouds and
clouds$+$lands, respectively (see Figure \ref{fig:fig6}a and
Figure \ref{fig:fig7}). Thus the difference of the photon counts
$N_1(t)$ and $N_4(t)$ roughly removes the contribution from clouds.

\begin{figure}
 \centering \includegraphics[width=16cm]
 {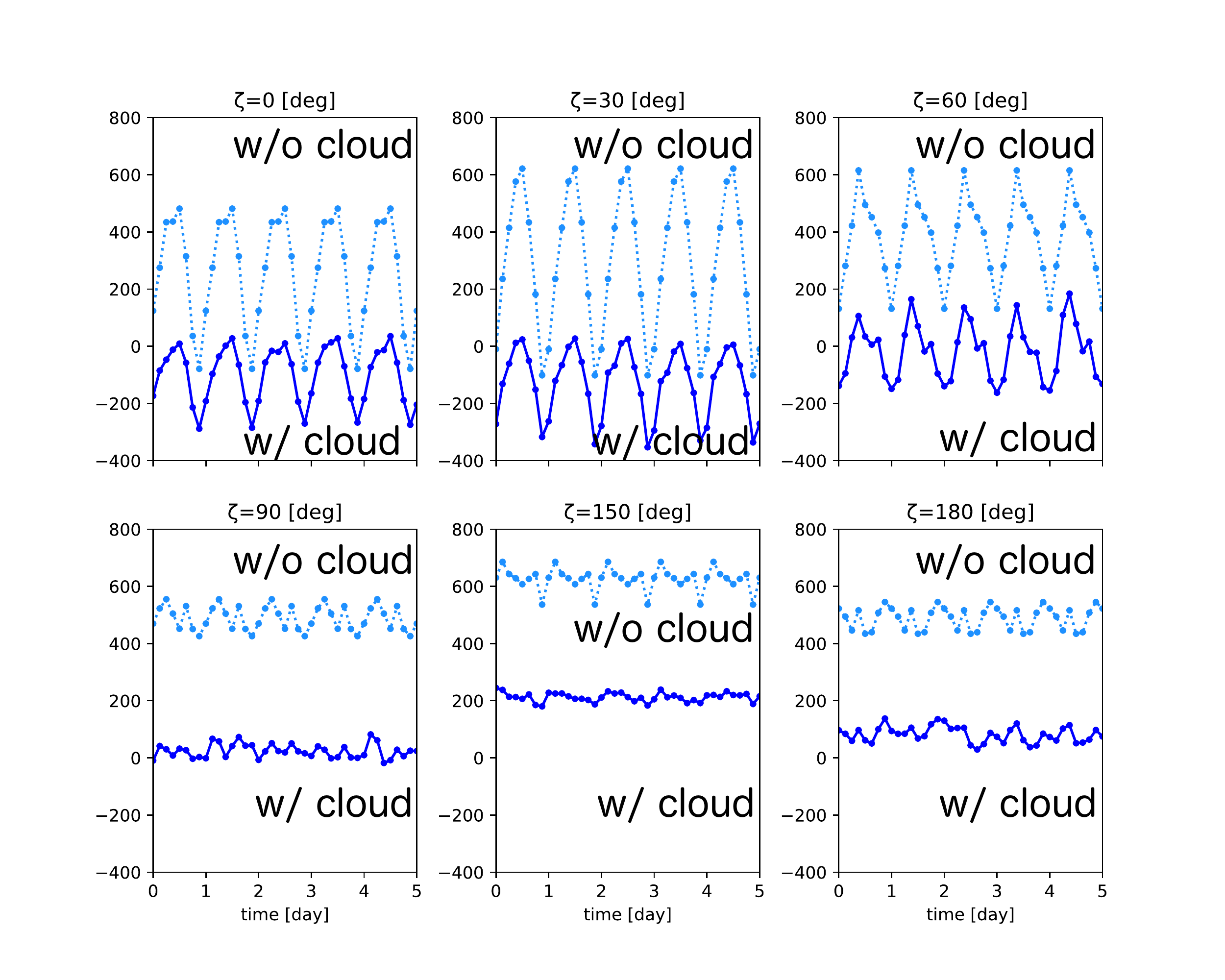}
  \caption{Simulated noiseless $C_{1-4}$ lightcurves for different
    obliquities ($\zeta=0^\circ$, $30^\circ$, $60^\circ$, $90^\circ$,
    $150^\circ$, and $180^\circ$). We adopt the same set of parameters
    as in Figure \ref{fig:fig9}.
\label{fig:fig13}}
\end{figure}

For definiteness, we choose the following
linear combination of bands 1 and 4:
\begin{eqnarray}
\label{eq:C14}
C_{1-4}(t) &=& N_1(t)-\alpha_{1-4}N_4(t), \\
\alpha_{1-4} &=& \frac{F_{*1}\lambda_1 \Delta\lambda}
{F_{*4}\lambda_4 \Delta\lambda}\sim 1.12 .
\end{eqnarray}
The above combination is derived assuming that the albedo of clouds is
roughly independent of the wavelength, and thus the contribution of
the clouds is canceled, at least partially as shown in Figure
\ref{fig:fig13}.  While the cloud effect may be removed more
efficiently by combining other
bands appropriately, it is beyond the scope of the present paper.
Thus we perform the frequency modulation analysis using
equation (\ref{eq:C14}) in what follows.

\begin{figure}[ht!]
\centering\includegraphics[width=16cm]{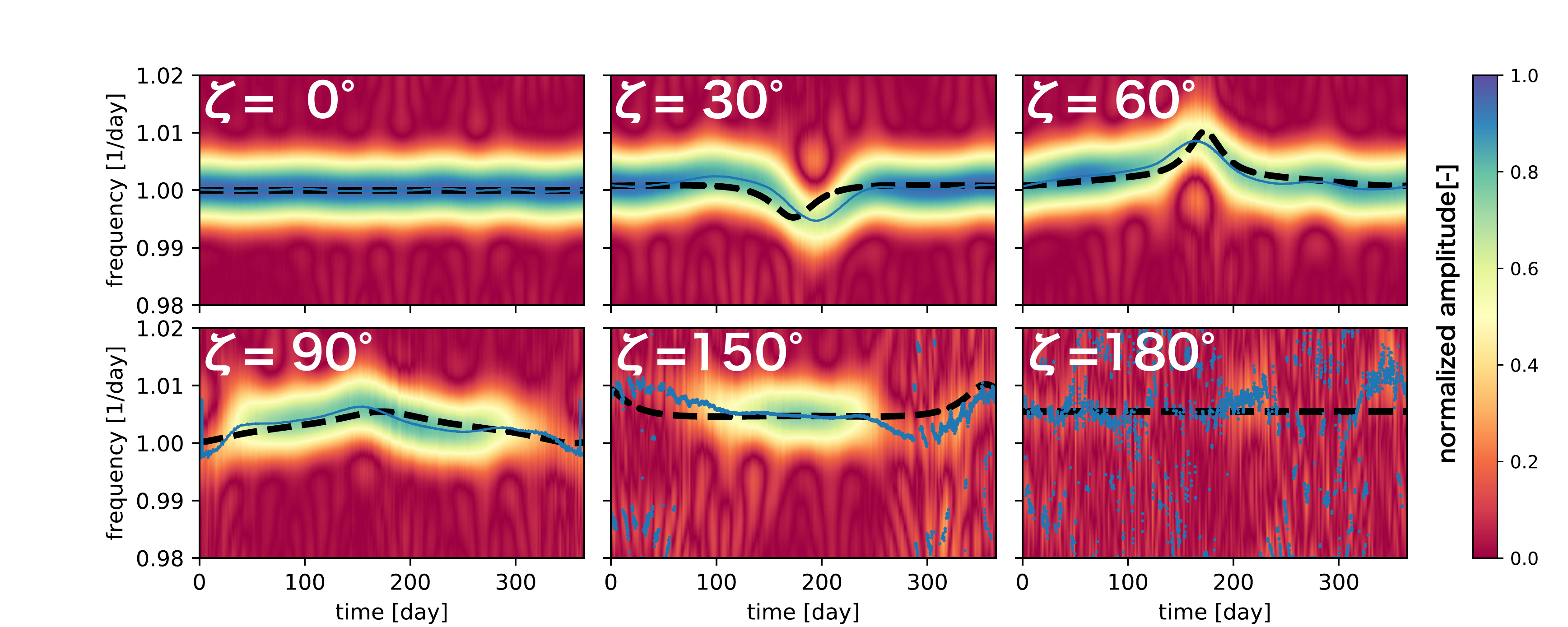}
\caption{The pseudo-Wigner distribution $g(f,t)$ of the noiseless
  $C_{1-4}$. Thick dashed lines indicate $f_{\rm model}(t)$, while
  thin blue points indicate $f_{\rm data, max}(t)$, the frequency
  corresponding to the maximum value of $g(f,t)$ over a range of
  $f_{\rm min}<f<f_{\rm max}$ at each epoch; see equation
  (\ref{eq:R1}). Due to the quality of the data, the values of $f_{\rm
    data, max}(t)$ are not robust for $\zeta=150^\circ$ and
  $180^\circ$, and discontinuous.
\label{fig:fig14}}
\end{figure}

Figure \ref{fig:fig13} shows an example of simulated
noiseless lightcurves using $C_{1-4}(t)$, and Figure
\ref{fig:fig14} is the corresponding time-frequency
representation.  Comparison between Figures \ref{fig:fig12} and
\ref{fig:fig14} clearly indicates that the multi-band
analysis suppresses the time-dependent cloud effect, and significantly
improves the frequency modulation signal.

We note that the amplitude of the frequency modulation signature
depicted in Figure \ref{fig:fig14} sensitively depends on the value of
$\zeta$, reflecting the specific surface distribution on the Earth.
As indicated in Figures \ref{fig:fig8a} and \ref{fig:fig8b}, the
Southern Hemisphere, especially around the South Pole, of the Earth is
occupied by Antarctica and ocean, in an approximately axisymmetric
manner.  Thus the diurnal variation of the Southern Hemisphere (for
example, viewed from the direction of $i=0^\circ$ if
$\zeta=180^\circ$) is difficult to detect.  This also applies to
the $\zeta =150^\circ$ case in which the frequency modulation signal
is clear only in summer, as described at the end of subsection
\ref{subsec:singleband}.

In contrast, the Northern Hemisphere is roughly divided into two major
distinct components; the Eurasian continent and the Pacific
ocean. This large-scale inhomogeneity, in particular the Sahara
desert, acts as a good tracer of an asymmetric surface pattern,
yielding a relatively large amplitude signal of the frequency
modulation (see Figures \ref{fig:fig8a} and \ref{fig:fig8b}).  This is
why a clear frequency modulation signal in the case of $\zeta \le
90^\circ$ can be detected for an observer located at $i=0^\circ$.

\subsection{Feasibility of the obliquity estimate \label{subsec:feasibility}}

\begin{figure*}
\centering\includegraphics[width=16cm]{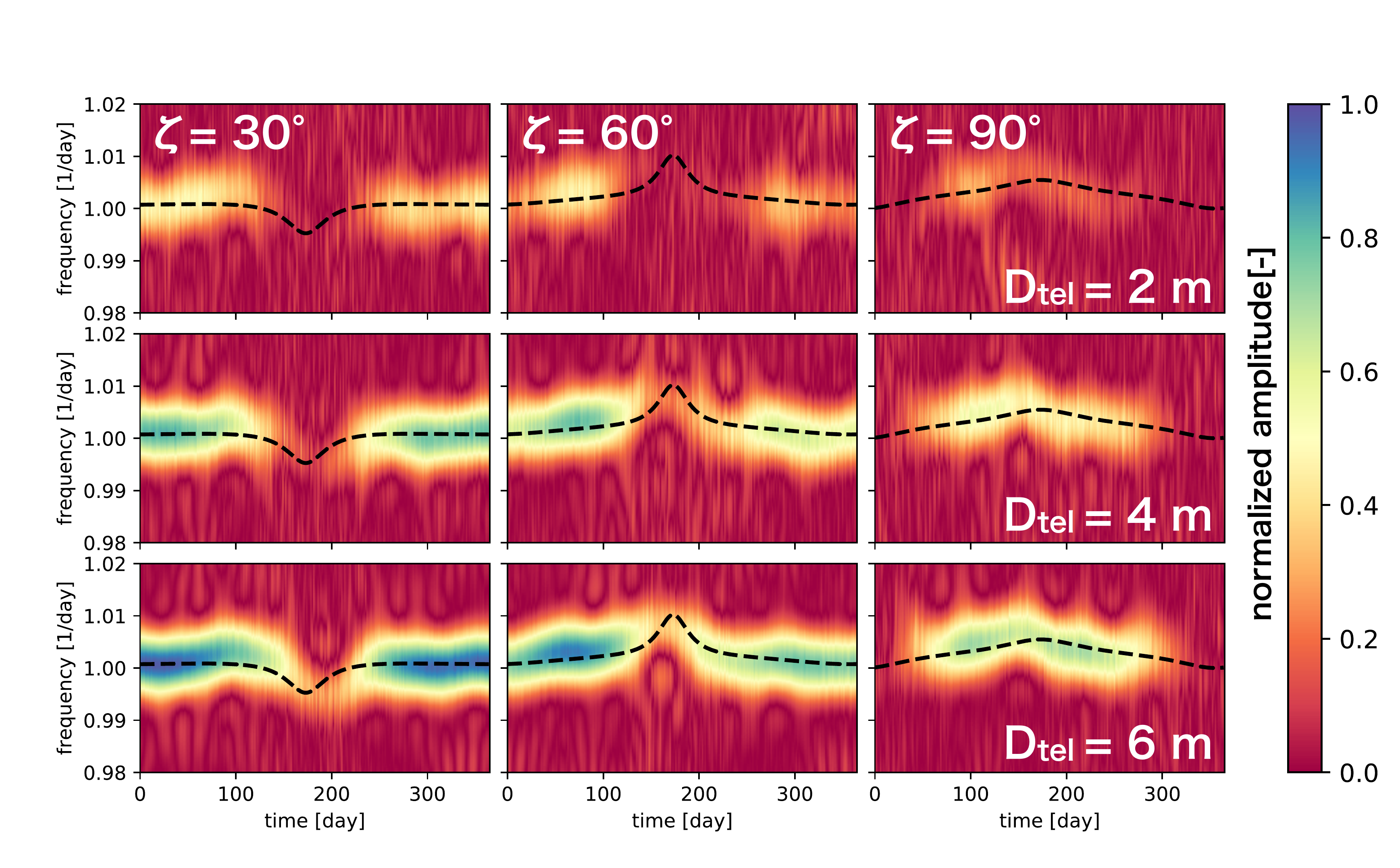}
\caption{The pseudo-Wigner distribution for oblique Earth-twins from
  the shot-noise limited photometric monitor. The top, middle, and
  bottom panels are for the space telescope aperture $D_{\rm tel}=2$,
  4, and 6 m, and the left, center, and right panels for the planetary
  obliquity $\zeta=30^\circ$, $60^\circ$ and $90^\circ$.  We assume
  that $D_{\rm obs}=10\;{\rm pc}$, $t_{\rm exp}=3\;{\rm hr}$,
  and $\Delta\lambda=0.1\,\mu{\rm m}$. \label{fig:fig15}}
\end{figure*}

All of the pseudo-Wigner distributions above (Figures \ref{fig:fig11},
\ref{fig:fig12}, and \ref{fig:fig14}) are based on noiseless data.
Now we are in a position to examine to what extent one can estimate
the planetary obliquity from the long-term photometric monitoring via
the frequency modulation method. For that purpose, we assume a
dedicated space mission with the telescope aperture of $D_{\rm
  tel}=2$, 4 and 6 m. Again we consider idealized cases in which the
photometric noise comes from the photon shot-noise alone, and generate
a set of $C_{1-4}(t)$ lightcurves from the photon counts $N_1(t)$ and
$N_4(t)$ obeying the Poisson statistics.  Examples of the resulting
frequency modulation are presented in Figure \ref{fig:fig15}.

The model frequency modulation is determined by the five parameters
($\zeta$, $f_{\rm spin}$, $\Theta_{\rm eq}$, $i$, $f_{\rm orb}$) that
are listed in Table \ref{tab:fit}; the planetary obliquity $\zeta$,
the planetary spin frequency $f_{\rm spin}$, the angle of the vernal
equinox measured from the location of the observer projected on the
orbital plane $\Theta_{\rm eq}$, the observer's inclination $i$, and
the orbital frequency of the planet $f_{\rm orb}$. Among them, $i$ and
$\Theta_{\rm eq}$ simply specify the location of the observer relative
to the system, and are not so interesting. The remaining three
parameters, $\zeta$, $f_{\rm spin}$ and $f_{\rm orb}$, are important
since they characterize the star-planet system.

In order to estimate $\zeta$, which cannot be estimated otherwise and
thus are of our primary interest, we need to perform eventually a
joint analysis of the five parameters in a Bayesian fashion.  In the
present study, however, we would like to examine the feasibility of
the determination of $\zeta$ and $f_{\rm spin}$, assuming that $i$ and
$f_{\rm orb}$ are known, for simplicity.  The precise spectroscopic
and astrometric data would determine $i$ and $f_{\rm orb}$. Also,
$f_{\rm spin}$ may be estimated from the photometric data on
relatively short-time scales apart from the uncertainty of
$\epsilon_\zeta(\Theta)f_{\rm orb}$ in equation (\ref{eq:fmodel}).

Under the similar assumption, K16  attempted
to find the best-fit values for $\zeta$, $f_{\rm spin}$, and
$\Theta_{\rm eq}$ by minimizing
\begin{equation}
\label{eq:R1}
R_1(\Theta_{\rm eq},\zeta, f_{\rm spin})
= \sum_{j=1}^{\rm N_{data}} \left|f_{\rm data, max}(t_j)
-f_{\rm model}(t_j;\Theta_{\rm eq},\zeta, f_{\rm spin}) \right|^2 ,
\end{equation}
where $f_{\rm model}(t)$ is the frequency derived from the {\it
  maximum-weighted longitude approximation}, equation
(\ref{eq:fmodel}), and $f_{\rm data, max}(t)$ corresponds to the
maximum value of $g(f,t)$ over a range of $f_{\rm min}<f<f_{\rm max}$
at each epoch $t$.  We tried the same fitting, but the result is not
robust against the shot noise especially when the frequency modulation
signal is weak.

Therefore we empirically improve the fit by taking account of the
distribution around the $f_{\rm data, max}(t)$ as well.  More
specifically, we construct a Gaussian weighted model
$\tilde{g}_{model}(f,t)$ for the time-frequency distribution:
\begin{equation}
\label{eq:tildeg-model}
\tilde{g}_{model}(f,t) = 
\exp\Bigg[-\frac{(f-f_{model}(t;\Theta_{\rm eq},\zeta, f_{\rm spin}))^2}
{2\sigma_f^2}\Bigg],
\end{equation}
where $\sigma_f$ is a new fitting parameter that is introduced to
account for the finite width of the frequency distribution
around $f_{\rm model}$.

Then we minimize the following quantity:
\begin{equation}
\label{eq:R2}
R_2(\Theta_{\rm eq},\zeta, f_{\rm spin},\sigma_f)
= \sum_{i=1}^{{\rm N}_f}\sum_{j=1}^{\rm N_{data}}
\left| \frac{g_{\rm data}(f_i,t_j)}{g_{\rm data}(f_{\rm data, max}(t_j),t_j)}
-\tilde{g}_{\rm model}(f_i,t_j) \right|^2,
\end{equation}
to find the best-fit $\zeta$, $\Theta_{\rm eq}$, $f_{\rm spin}$, and
$\sigma_f$. The value of $\sigma_f$ should be roughly equal to
$1/T_{\rm w}$ because the time-frequency representation of a
signal $z(t)=e^{if_{\rm ins}t}$ based on the pseudo-Wigner
distribution has a dispersion corresponding to the Fourier transform
of the window function $\tilde{h}(f-f_{\rm ins}; T_w)$, and this
dispersion is flattened due to the noise and non-linear frequency
modulation.

In practice, we use the Levenberg-Marquardt algorithm
{\tt mpfit}\citep{2009ASPC..411..251M} to find the best-fit parameters.
This algorithm is a practical and fast algorithm of the least square
method for non-linear functions.  We fit the time-frequency
distribution for $\zeta=30^\circ$ and $60^\circ$.  Table
\ref{tab:fit} summarizes our initial parameters in addition to the
fixed orbital parameters that we assume to be {\it a priori} known.

\begin{table}[htb]
 \caption{Initial and fixed parameter sets for best fit search}
  \centering
   \begin{tabular}{|l|c|} \hline
Initial parameter & value\\ \hline \hline
obliquity $\zeta$ & 15$i^\circ$ ($i=1,2,\cdots,12$)\\
$\Theta_{\rm eq}$ & $60j^\circ$ ($j=1,2,\cdots,5$)\\
spin frequency $f_{\rm spin}$ & 366 [year$^{-1}$]\\ \hline \hline
Fixed parameter & value\\ \hline \hline
orbital inclination $i$ & 0$^\circ$\\
orbital frequency $f_{\rm orb}$ & 1 [year$^{-1}$]\\ \hline
   \end{tabular}
\label{tab:fit}
\end{table}

\begin{figure*}
 \centering \includegraphics[width=14cm]{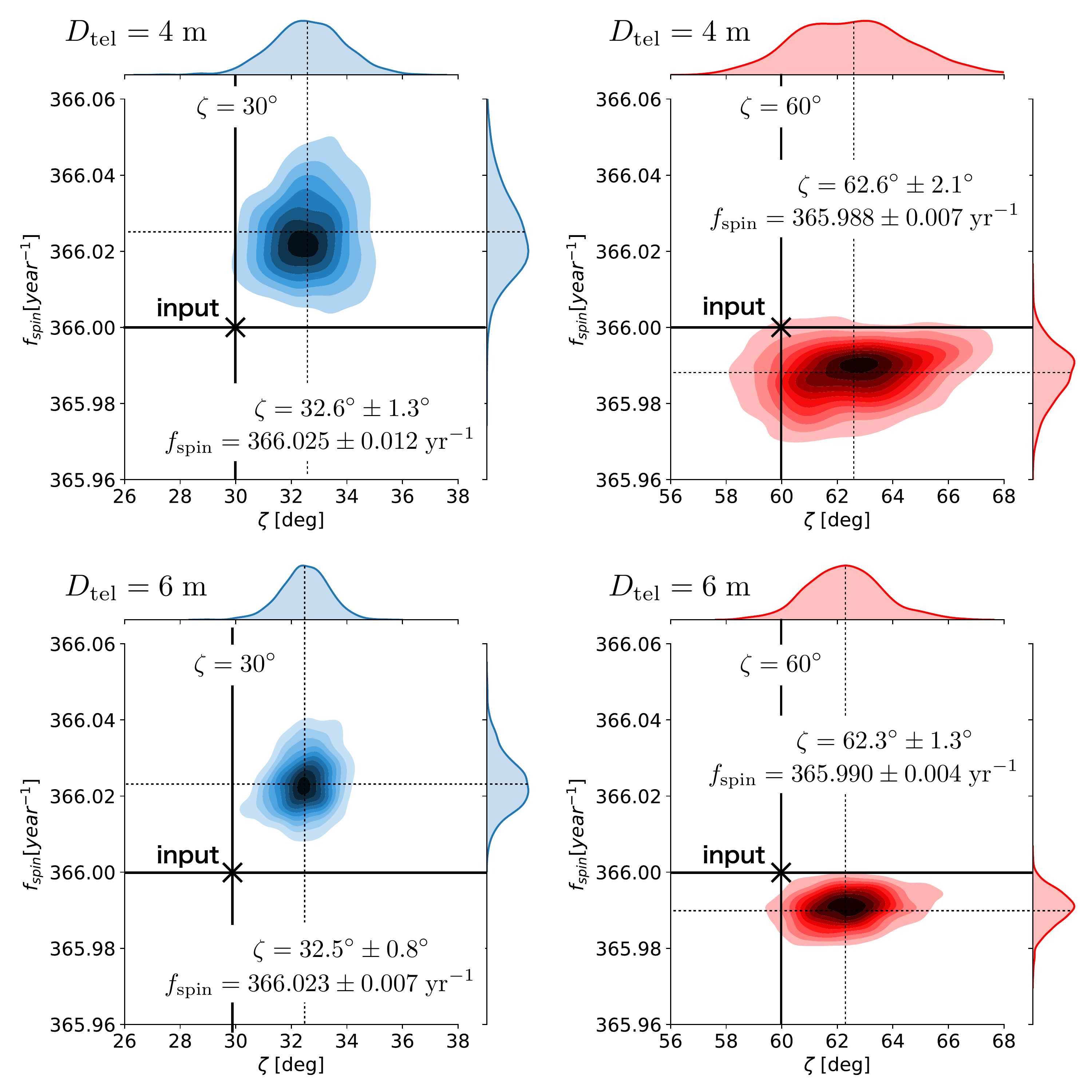}
\caption{Spin parameters ($\zeta$, $f_{\rm spin}$) estimated from the
  normalized Gaussian model $\tilde{g}_{model}$. Black crosses show
  the input spin values ($\zeta$, $f_{\rm spin}$) = (30$^\circ$, 366
  year$^{-1}$) and (60$^\circ$, 366 year$^{-1}$) for left and right
  panels, respectively. We plot the best fit values from 1000
  realizations of shot-noise limited observation. The shot noise is
  assumed from the telescope diameter of $D_{\rm tel} = 4$ m and
    6 m for top and bottom panels, respectively.}
\label{fig:fig16}
\end{figure*}

Figure \ref{fig:fig16} shows the distribution of the best-fit
estimates on the $\zeta$-$f_{\rm spin}$ plane from 1000 different
realizations. The black cross symbols indicate the input values,
($\zeta,$ $f_{\rm spin}$) = ($30^\circ$, 366 year$^{-1}$) and
($60^\circ$, 366 year$^{-1}$), for left and right panels,
respectively. The top and bottom panels show the results based
on the shot-noise limited observations with $D_{\rm tel}=$ 4 m and 6 m,
respectively. The numbers in each panel
denote the mean and 1$\sigma$ estimated from 1000 realizations.

The systematic offsets of $(\Delta\zeta)_{\rm sys} \approx 3^\circ$
and $(\Delta f_{\rm spin})_{\rm sys}\approx 0.03$ year$^{-1}$ result
most likely from the specific pattern of the continents on the
Earth. Indeed the previous simplified analysis by K16 also found a
similar level of the systematic offset of the planetary obliquity
($\sim$ several degrees; see Figure 8 of K16). K16 added noises
empirically into his mock data, neglecting the time-dependent cloud
distribution that we compute here.

The fact that the systematic offsets between the two analyses are
similar indicates, therefore, that they should be ascribed to the
specific surface pattern of the Earth itself.  Indeed Eurasia, North
Africa, and South America are distributed roughly from northeast to
southwest directions. This latitudinal pattern is consistent with the
positive systematic offset of the obliquity exhibited in Figure
\ref{fig:fig16}. Since the amplitude of the systematic offset would
depend on the specific pattern of the planetary surface to some
extent, it is difficult to predict it {\it a prior}, but important to
bear in mind that it could amount to several degrees, much larger than
the statistical uncertainty as shown in Figure \ref{fig:fig16}.

\section{Summary and conclusion \label{sec:conclusion}}

The direct imaging of earth-like planets is very challenging, but will
provide ground-breaking datasets for astronomy, planetary science, and
biology, if successful eventually.  One notable example is the
reconstruction of the surface components
\citep[e.g.,][]{1993Natur.365..715S, 2001Natur.412..885F,
  2010ApJ...715..866F, 2011ApJ...738..184F, 2011ApJ...739L..62K,
  2012ApJ...755..101F,2019asbi.book..441S}, and it may be even
possible to measure the planetary obliquity through the frequency
modulation of the photometric lightcurve of future direct imaged
Earth-like planets, proposed by \citet{2016ApJ...822..112K}.

We have examined the feasibility of the methodology by creating
simulated lightcurves of our Earth-Sun systems but with different
planetary obliquities. First, we performed the GCM simulation for
those systems with particular emphasis on the time-dependent cloud
distribution. Second, we computed the scattered light in 6 photometric
bands by solving the radiation transfer of the incident starlight
through the cloud and atmosphere taking into account the scattering
due to the different surface components under the parameterized
bi-directional reflectance distribution function models
\citep{JGRD:JGRD3769,1983JQSRT..29..521N}. Third, the resulting light
from the planet was mock-observed every three hours over the orbital
period of one year, and simulated lightcurves were constructed by
combining the different photometric bands so as to suppress the effect
of the time-dependent cloud pattern. Finally, we computed the
frequency modulation of the lightcurves using the pseudo-Wigner
distribution and attempted to estimate the planetary
obliquities for photon-shot noise dominated cases.

We found that the frequency modulation signal is crucially dependent
on the presence of the large-scale inhomogeneity on the planetary
surface. Indeed this is the case for the Northern Hemisphere of our
Earth; in particular the Sahara desert turned out to be a useful
tracer of the planetary spin rotation. The Southern Hemisphere, on the
other hand, is relatively featureless, and the frequency modulation
signal is weak.

As a result, we found that a dedicated 4 m space telescope at 10 pc
away from the system in the face-on view relative to the
observer can estimate the planetary obliquity within the
uncertainty of several degrees in principle (in the shot-noise limited
case).  Although this conclusion is based on several idealized
assumptions at this point, we believe that it is very encouraging for
the future exploration of the direct imaging of Earth-like planets.

\acknowledgements

We thank an anonymous referee for numerous constructive suggestions
and comments that significantly improved the early manuscript of the
paper. This work is supported by Japan Society for the Promotion of
Science (JSPS) Core-to-Core Program “International Network of
Planetary Sciences”, the European Research Council under the European
Union's Horizon 2020 research and innovation programme (Grant
Agreement 679030/WHIPLASH), JSPS KAKENHI Grant Numbers JP18H01247 and
JP19H01947 (Y.S.), JP17K14246 and JP18H04577 (H.K.), JP19H01947
(M.I.), JP17H06457 (K.K. and Y.O.T), and the Astrobiology Center
Program of National Institutes of Natural Sciences (Grant Number
AB311025).  Numerical computations and analyses were partly carried
out on PC cluster at Center for Computational Astrophysics, National
Astronomical Observatory of Japan.  Y.N. acknowledges the support by
fellowship from the Advanced Leading Graduate Course for Photon
Science at the University of Tokyo.

  \software{
    {\tt DCPAM5} ({\tt http://www.gfd-dennou.org/library/dcpam}), 
    {\tt libRadtran}\citep{gmd-9-1647-2016,acp-5-1855-2005},
    {\tt REPTRAN}\citep{2014JQSRT.148...99G}, 
    {\tt mpfit}\citep{2009ASPC..411..251M},
{\tt juwvid} ({\tt https://github.com/HajimeKawahara/juwvid}).
  }

\begin{appendix}
\section{Behavior of the frequency modulation factor
$\epsilon_\zeta(\Theta)$ for an eccentric orbit \label{sec:epsilon}}

The modulation factor, equation (\ref{eq:emodel}), first derived by
K16 assumes a circular orbit for simplicity.  We compute a generalized
expression for an eccentric orbit, and present the effect of
eccentricity on the frequency modulation based on the {\it maximum
  weighted longitude approximation}.

\begin{figure}[htbp]
\centering\includegraphics[width=12cm]{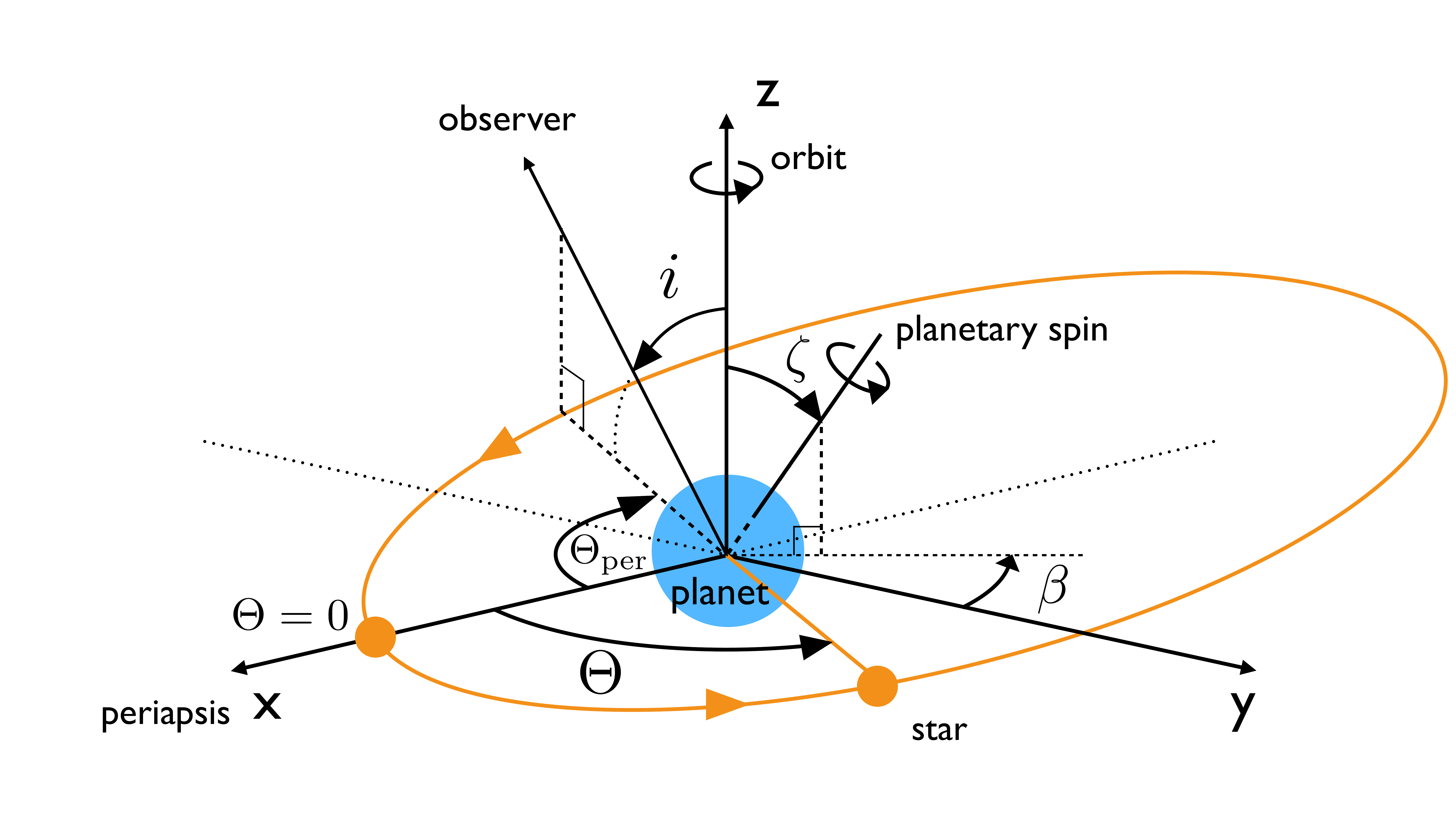}
\caption{A schematic configuration of a
geocentric frame for eccentric orbits.
 \label{fig:eccentric-frame}}
\end{figure}

For an eccentric orbit, it is more convenient to consider a geocentric
frame where the $x$-axis is the direction toward the periapsis as
shown in Figure \ref{fig:eccentric-frame}, instead of vernal equinox
({\it c.f.,} Figure \ref{fig:fig3}). In this frame, the spin vector
is no longer on the $yz$-plane, and we introduce a new parameter
$\beta$, which denotes the azimuthal angle of the planetary spin
measured from $y$-axis. Similarly the location of the observer is
specified by the phase angle from the periapsis $\Theta_{\rm per}$,
and $\Theta$ is now the azimuthal angle measured clockwise from the
periapsis, {\it i.e.}, the true anomaly.  The frame reduces to that
shown in Figure \ref{fig:fig3} for $e \to 0$, $\beta \to 0^\circ$,
$\Theta_{\rm per} \to \Theta_{\rm eq}$, and $\Theta \to
\Theta-\Theta_{\rm eq}$.

Following K16, we assume that the longitude of the reflective point on
the planetary surface, $\hat{\phi}_{\rm M}$, traces faithfully the
observable periodicity of the planetary scattered light. Then the
observed frequency $f_{\rm obs}$ is given in terms of $\hat{\phi}_{\rm
  M}$ as
\begin{eqnarray}
\label{eq:eccentric-fobs}
f_{\rm obs}(t) &=& -\frac{1}{2\pi}\frac{{\rm d} \hat{\phi}_{\rm M}}{{\rm d} t}
= -\frac{1}{2\pi}\frac{{\rm d} \Theta}{{\rm d}  t}
\frac{\partial \hat{\phi}_{\rm M}}{\partial \Theta} \cr
&=& f_{\rm spin} + \frac{1}{2\pi}\frac{{\rm d} \Theta}{{\rm d}  t}
\epsilon_\zeta(\Theta),
\end{eqnarray}
where 
\begin{eqnarray}
\label{eq:eccentric-epsilon}
\epsilon_\zeta(\Theta)
& \equiv & - \frac{\partial (\hat{\phi}_{\rm M}+\Phi)}{\partial \Theta}
= - \frac{\kappa '(\Theta)}{1+\kappa(\Theta)^2}, \\
\label{eq:eccentric-kappa}
\kappa(\Theta)
& \equiv & \tan(\hat{\phi}_{\rm M}+\Phi), \\
\label{eq:eccentric-Phi}
\Phi & \equiv & 2\pi f_{\rm spin} t. 
\end{eqnarray}
For a circular orbit, $d\Theta/dt$ is equal to $f_{\rm orb}$.  For $e
\not=0$, however, it cannot be written explicitly in terms of $t$, but
is expressed as
\begin{equation}
\label{eq:eccentric-dtheta-dt}
\frac{d\Theta}{dt}
=  2\pi f_{\rm orb}\;(1-e^2)^{-3/2}(1+e\cos\Theta )^2.
\end{equation}

In the geocentric frame, unit vectors toward the star and the
observer, $\vec{e}_s$ and $\vec{e}_o$, are given as $\vec{e}_s=(\cos
\Theta, \sin \Theta, 0)$, and $\vec{e}_o = (\cos \Theta_{\rm per}\sin
i, -\sin \Theta_{\rm per}\sin i, \cos i)$, respectively. Thus the unit
vector towards the reflective point, $\vec{e}_{\rm M}$, is
\begin{eqnarray}
\label{eq:eccentric-vec-M}
\vec{e}_{\rm M}
=  \frac{\vec{e}_s+\vec{e}_o}{|\vec{e}_s+\vec{e}_o|} 
= 
\frac{1}{L}\left(
\begin{array}{c}
\cos \Theta + \cos \Theta_{\rm per}\sin i\\
\sin \Theta - \sin \Theta_{\rm per}\sin i\\
\cos i
\end{array}     \right) ,
\end{eqnarray}
where $L\equiv|\vec{e}_s+\vec{e}_o|=\sqrt{2+2\cos(\Theta+\Theta_{\rm
    per})\sin i}$.

Consider a point on the planetary surface specified by the latitude
$\lambda$ and longitude $\phi$ in the rest frame of the planet.  The
surface normal unit vector at the point is $\vec{e_R}'(\phi, \lambda)
= (\cos\phi\cos\lambda, \sin\phi\cos\lambda, \sin\lambda)$. One can
transform $\vec{e_R}'$ to $\vec{e_R}$ in the geocentric frame as
\begin{equation}
\label{eq:eccentric-transform-forward}
\vec{e}_R = R_z(\beta) R_x(-\zeta) \hat{\rm S}(\Phi) \vec{e_R}', 
\end{equation}
where $\hat{\rm S}(\Phi)$ is a spin rotation operator
($\phi\to\phi+\Phi$), and $R_i$ is the rotation matrix
counterclockwise around the $i$-axis. Note that $R_z(\beta)$ is
required for a non-circular case.

\begin{figure*}
\gridline{
        \fig{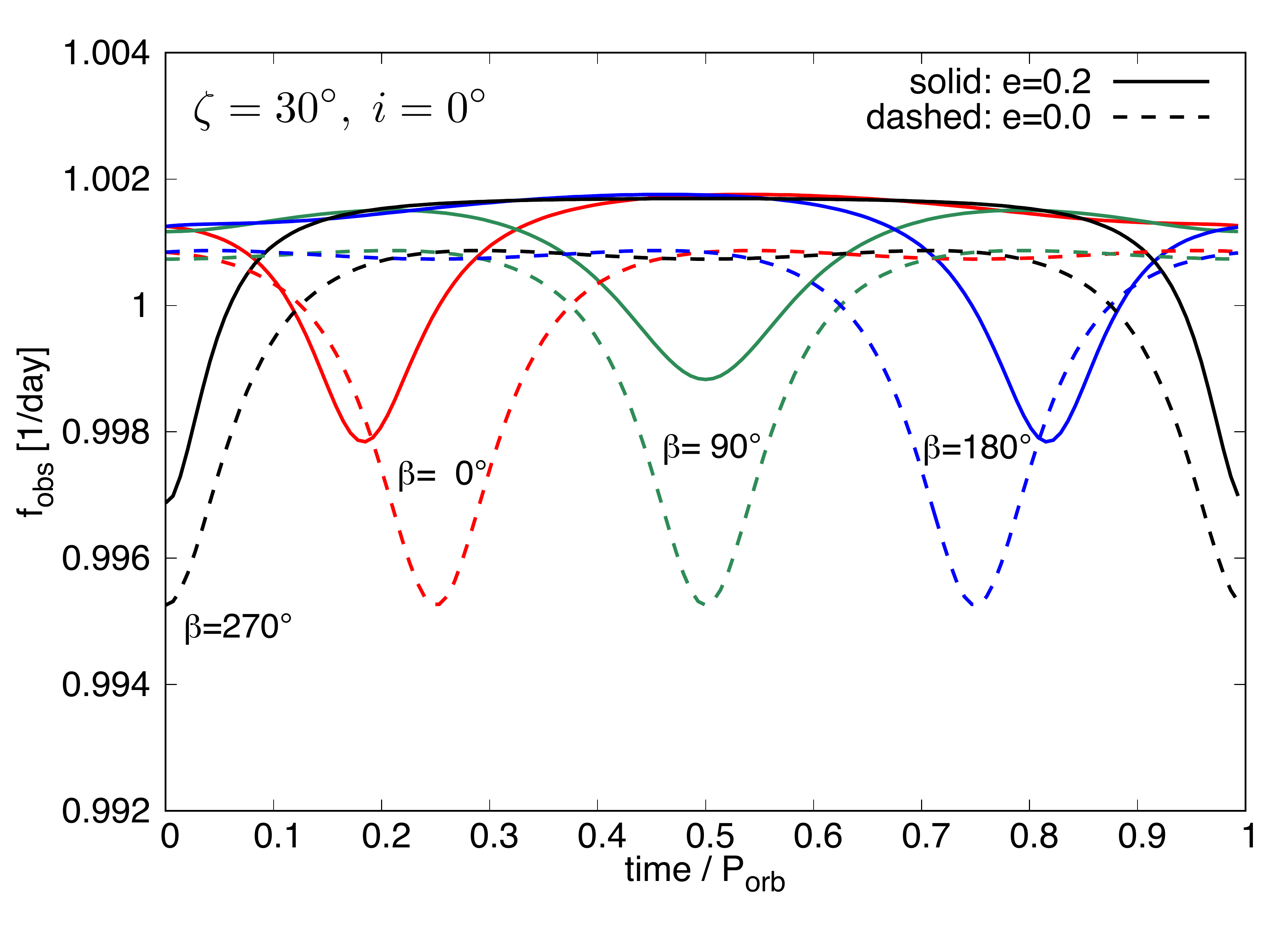}
        {0.5\textwidth}
        {(a) }
        \fig{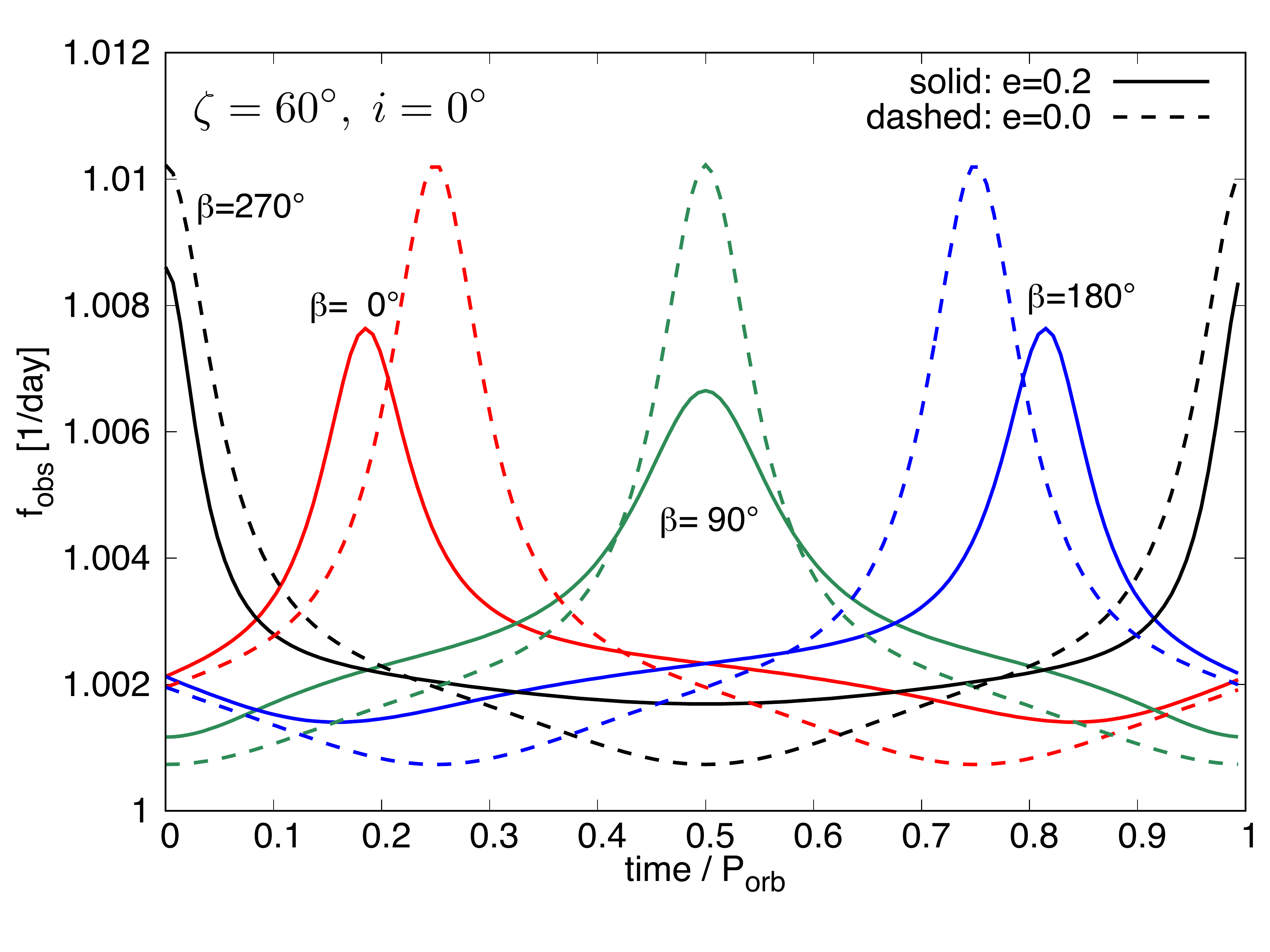}
        {0.5\textwidth}
        {(b)}
        }
\caption{Samples of frequency modulation applied to eccentric orbits;
 (a) $\zeta=30^\circ$ and (b) $\zeta=60^\circ$. Each line shows the
 frequency modulation specific to the spin configuration of the
 planet. Solid lines are eccentric cases ($e=0.2$) and dashed lines
 are circular cases ($e=0.0$).  \label{fig:eccentric-FM}}
\end{figure*}

We apply the generic transformation, equation
(\ref{eq:eccentric-transform-forward}), to compute the component of
the reflective point in the planetary frame.  Then we obtain
\begin{eqnarray}
\label{eq:eccentric-transform-backward}
\vec{e_{\rm M}}'(\hat{\phi}_{\rm M}+\Phi) 
&=& \hat{\rm S}(\Phi) \vec{e_{\rm M}}'(\hat{\phi}_{\rm M}) 
=  R_x(\zeta) R_z(-\beta) \vec{e_{\rm M}} \cr
& = & \frac{1}{L}\left(
\begin{array}{c}
\cos(\Theta-\beta)+\sin i \cos(\Theta_{\rm per} + \beta)\\
\cos\zeta\{\sin(\Theta-\beta)-\sin i \sin(\Theta_{\rm per} + \beta)\}
-\sin\zeta\cos i\\
\sin\zeta\{\sin(\Theta-\beta)-\sin i \sin(\Theta_{\rm per} + \beta)\}
+\cos\zeta\cos i
\end{array}     \right) .
\end{eqnarray}
The ratio of the $x$- and $y$-components in equation
(\ref{eq:eccentric-transform-backward}) yields
\begin{equation}
\label{eq:eccentric-kappa-formula}
\tan(\hat{\phi}_{\rm M}+\Phi)=
\frac{\cos\zeta\{\sin(\Theta-\beta)-\sin i \sin(\Theta_{\rm per} + \beta)\}
-\sin\zeta\cos i}
{\cos(\Theta-\beta)+\sin i \cos(\Theta_{\rm per} + \beta)}.
\end{equation}
Therefore equation (\ref{eq:eccentric-epsilon}) reduces to
\begin{equation}
\label{eq:eccentric-epsilon-formula}
\epsilon_\zeta(\Theta)=
\frac{-\cos\zeta\{1+\sin i \cos(\Theta+\Theta_{\rm per})\}
  +\sin\zeta\cos i\sin(\Theta - \beta) }
{\Bigl[\cos(\Theta-\beta)+\sin i \cos(\Theta_{\rm per} + \beta)\Bigr]^2
+\Bigl[\cos\zeta\{\sin(\Theta-\beta)
-\sin i \sin(\Theta_{\rm per} + \beta)\} -\sin\zeta\cos i\Bigr]^2}.
\end{equation}

Finally we obtain the eccentric frequency modulation in terms of true
anomaly $\Theta$:
\begin{equation}
\label{eq:eccentric-fobs-formula}
f_{\rm obs} = f_{\rm spin} + 
\frac{f_{\rm orb}\;(1-e^2)^{-3/2}(1+e\cos\Theta )^2
  \Bigl[-\cos\zeta\{1+\sin i \cos(\Theta+\Theta_{\rm per})\}
    +\sin\zeta\cos i\sin(\Theta - \beta)\Bigr] }
     {\Bigl[\cos(\Theta-\beta)+\sin i \cos(\Theta_{\rm per} + \beta)\Bigr]^2
       +\Bigl[\cos\zeta\{\sin(\Theta-\beta)
         -\sin i \sin(\Theta_{\rm per} + \beta)\} -\sin\zeta\cos i\Bigr]^2}. 
\end{equation}
The above equation reproduces equation (\ref{eq:emodel}) for $e \to
0$, $\beta \to 0^\circ$, $\Theta_{\rm per} \to \Theta_{\rm eq}$, and
$\Theta \to \Theta-\Theta_{\rm eq}$.

In the main part of the paper, we consider the circular orbit alone
just for simplicity, but the effect of $e$ is important as well.  To
show this, we plot equation \ref{eq:eccentric-fobs-formula} for the
Earth-like planet viewed from $i=0^\circ$ for $e=0$ and $0.2$ in
Figure \ref{fig:eccentric-FM}.  The horizontal axis indicates the time
in units of $P_{\rm orb}$ that is numerically computed from the true
anomaly. The left and right panels correspond to the obliquity of
$\zeta=30^\circ$ and $\zeta=60^\circ$.  Different curves indicate
cases for $\beta=0^\circ$, $90^\circ$, $180^\circ$, and $270^\circ$.

The dashed ($e=0$) and solid ($e=0.2$) clearly exhibit different
amplitudes and phases of the frequency modulation. Thus the effect of
eccentricity biases the estimates of $\zeta$ and $\Theta_{\rm eq}$ if
the formula for $e=0$ is used in fitting the data.  Since it is likely
that the orbit of direct imaging targets is precisely determined prior
to its monitoring, one can use equation
(\ref{eq:eccentric-fobs-formula}) as a template using the estimated
value of $e$.

\end{appendix}



\end{document}